\documentclass[authoryear,12pt]{article}%
\usepackage{color}
\usepackage{amsfonts}
\usepackage{amsthm}
\usepackage{amssymb,bm}
\usepackage{amsmath,enumerate,rotate}
\usepackage{amssymb,latexsym,mathrsfs}
\usepackage{graphicx}
\usepackage[font=footnotesize]{caption}
\usepackage{amsmath}
\usepackage{amssymb}
\usepackage{sectsty}
\usepackage{setspace}
\usepackage{titlesec}
\usepackage[left=.58in,right=0.64in,top=1in,bottom=1in]{geometry}
\usepackage{url}
\usepackage{natbib}%
\providecommand{\U}[1]{\protect\rule{.1in}{.1in}}
\RequirePackage[OT1]{fontenc}
\RequirePackage{amsthm,amsmath}
\RequirePackage[colorlinks,citecolor=blue,urlcolor=blue]{hyperref}
\numberwithin{equation}{section}
\theoremstyle{plain}
\newtheorem{thm}{Theorem}[section]

\theoremstyle{definition}

\theoremstyle{remark}

\sectionfont{\normalsize}
\subsectionfont{\normalsize\normalfont\itshape}
\titlelabel{\thetitle.\quad}

\subsubsectionfont{\normalsize\normalfont\itshape}
\begin{document}

\title{\singlespacing\centering{\large \textbf{Bayes Calculations From Quantile
Implied Likelihood}}}
\author{{\normalsize George Karabatsos}\\{\normalsize University of Illinois-Chicago}
\and {\normalsize and Fabrizio Leisen}\\{\normalsize University of Kent}}
\date{{\normalsize \today}}
\maketitle

\singlespacing\abstract{\small{\textbf{Abstract}
In statistical practice, a realistic Bayesian model for a given data set can be defined by a likelihood function that is analytically or computationally intractable, due to large data sample size, high parameter dimensionality, or complex likelihood functional form. This in turn poses challenges to the computation and inference of the posterior distribution of the model parameters. For such a model, a tractable likelihood function is introduced which approximates the exact likelihood through its quantile function. It is defined by an asymptotic chi-square confidence distribution for a pivotal quantity, which is generated by the asymptotic normal distribution of the sample quantiles given model parameters. This Quantile Implied Likelihood (QIL) gives rise to an approximate posterior distribution which can be estimated by using penalized log-likelihood maximization or any suitable Monte Carlo algorithm. The QIL approach to Bayesian Computation is illustrated through the Bayesian analysis of simulated and real data sets having sample sizes that reach the millions. The analyses involve various models for univariate or multivariate iid or non-iid data, with low or high parameter dimensionality, many of which are defined by intractable likelihoods. The probability models include the Student's t, g-and-h, and g-and-k distributions; the Bayesian logit regression model with many covariates; exponential random graph model, a doubly-intractable model for networks; the multivariate skew normal model, for robust inference of the inverse-covariance matrix when it is large relative to the sample size; and the Wallenius distribution model.}%
}
\newpage%
\setstretch{1}%

\section{Introduction}

For any Bayesian model, statistical inference focuses on the posterior
distribution of the model's parameters, which combines the model's data
likelihood with a prior distribution on the parameter space. However, a
realistic Bayesian model for the data may be defined by a likelihood function
that is intractable or difficult to manage, possibly due to large data sample
size, number of parameters, or complex likelihood functional form. Then,
computations of the posterior distribution can become slow, cumbersome, or
even impossible.

In such a situation, a likelihood-free method can provide a tractable and
surrogate approximate likelihood, which when combined with the chosen prior
distribution, yields an approximate posterior distribution that can be more
rapidly estimated using any suitable Markov chain Monte Carlo (MCMC),
importance sampling (IS), or other Monte Carlo (MC)\ or optimization-based
iterative computational algorithm
\citep[e.g.,][]{RobertCasella04}%
. Each likelihood-free method employs a specific approximate likelihood
function, constructed either by the rejection or kernel Approximate Bayesian
Computation (ABC), synthetic likelihood (SL), empirical likelihood (EL), or
bootstrap likelihood (BL)\ approach
\citep[][and references therein]{KarabatsosLeisen18abc}%
. However, these likelihood-free methods have certain limitations despite
their past successes. For such a method, posterior distribution estimates can
significantly depend on the analyst's choice of multiple tuning parameters
(ABC); or may involve, per algorithm iteration, a construction of the
approximate likelihood that is computationally-costly or cumbersome especially
when the data sample size is large. This construction may involve point
estimation of parameters (SL, EL, SL, BL), drawing one or more samples of
synthetic data sets from the likelihood (ABC, SL), or generating bootstrap
resamples of the original data set (BL)
\citep[][]{KarabatsosLeisen18abc}%
. Apparently, there is a current need to develop new likelihood-free method
that is defined by an approximate likelihood which is directly determined by
available statistical theory, does not rely on multiple tuning parameters, and
can be directly and rapidly computed in each iteration of the given posterior
distribution estimation algorithm.

To address this need, we introduce the Quantile Implied Likelihood (QIL). The
QIL is a novel approximate likelihood based on the asymptotic implied
likelihood approach
\citep{Efron93}
to confidence distribution theory
\citep[e.g.,][]{XieSingh13}%
, which we will show is useful for providing approximate posterior inference
for a wide range of Bayesian models with intractable likelihoods. The QIL is
an asymptotic chi-square ($\chi_{d}^{2}$) pdf that can be directly and
efficiently computed, and avoids many of the problems of the previous
likelihood-free methods. The QIL\ can have lower computational cost than the
synthetic, empirical, and bootstrap likelihoods, and does not rely on multiple
ABC tuning parameters, point-estimation, or synthetic data or bootstrap
sampling. Indeed, low computational cost is a hallmark of implied likelihoods
\citep{Efron93}%
.

The QIL is the asymptotic $\chi_{d}^{2}$\ pdf of the sampling distribution for
a pivotal quantity of a new quantile-based distribution test of the null
hypothesis that the true data-generating distribution equals the likelihood
distribution given the model parameters, against the alternative hypothesis of
inequality. This pivotal quantity is defined by the Mahalanobis distance
between data sample quantiles, and the quantiles of the model likelihood on
given parameters (which are directly computable or numerically approximated),
on $d\leq n$ evenly-spaced (cdf) probabilities and degrees of freedom.\ The
pivotal quantity is $\chi_{d}^{2}$ distributed under the null hypothesis by
virtue of the asymptotic multivariate normality of the $d$ sample quantiles
conditionally on the parameters. Univariate quantiles are naturally and
coherently defined for univariate data, and also for multivariate data; that
is, for the depths of multivariate observations (resp.)\
\citep{LiuSingh93}%
. In either case, the QIL may be efficiently computed from a number of $d\leq
n$ quantiles that can be chosen naturally to yield sample quantiles that are
close in Kolmogorov distance to the full data set, even for Big Data (large
$n$) sets, while still factoring in the total sample size $n$ in the
QIL\ likelihood computations. The QIL can be extended to non-iid grouped (or
regression) data using the same conditional independence assumptions as those
of the exact model likelihood.

For the given Bayesian model, the approximate posterior distribution is formed
by combining the QIL\ and the chosen prior distribution for the model
parameters. In addition, the QIL's simple and explicit $\chi^{2}$ pdf form
makes it possible to use with any standard MC or optimization algorithm for
posterior estimation, unlike the EL\ method which relies exclusively on IS
algorithms. Further, the maximum a posteriori (MAP) and posterior covariance
matrix estimates of the model parameters can be quickly estimated through
penalized QIL maximization, adding to previous work on maximum intractable
likelihood estimation
\citep{RubioJohansen13}%
.

More details about the QIL approach to Bayesian inference are described next
in $\S 2$. In $\S 3$, the QIL\ approach is illustrated on $27$ Bayesian models
through the analysis of many large or complex real and simulated data sets.
Results include posterior distribution estimates, computation times, and
accuracies of QIL-based posterior inferences for the simulated data in terms
of Root Mean Squared Error (RMSE).

The $27$ Bayesian models include $19$ standard low-dimensional parametric
probability models for univariate iid data, which provide basic benchmark
tests of QIL\ (e.g., Student's \textit{t} model). They also include models
with more intractable likelihoods, which are discussed in the literature and
summarized as follows:

\begin{enumerate}
\item \textbf{(Univariate iid data).} The generalized $g$-and-$h$ and
$g$-and-$k$ distributions
\citep{MacGillivray92}
are each defined by a likelihood function through its quantile function. This
likelihood has no closed-form expression and high computational cost for large
data sets. For each model, we will show that the QIL provides improved
accuracy in posterior inferences and competitive computational speed compared
to standard ABC methods.

\item \textbf{(Univariate non-iid data).} The Bayesian binary regression model
has regression coefficients that may be assigned a multivariate normal prior,
or even a LASSO prior with unknown shrinkage hyperparameter for variable
selection. For the binary logit or probit model, standard Gibbs sampling MCMC
algorithms are computationally slow for large data samples sizes because they
rely on iterative sampling of latent variables underlying the observed binary
dependent responses (resp.). Such an algorithm is further slowed when a
LASSO\ prior is assigned to the regression coefficients. This is because then,
per sampling iteration, additional steps are needed to perform sampling
updates of the coefficient scale parameters and the shrinkage hyperparameter,
and to perform a matrix inversion to obtain the conditional posterior
covariance matrix of the coefficients
\citep[][]{ParkCasella08}%
. We will show that the QIL, when serving as an approximate likelihood for the
Bayesian binary regression model, can lead to faster posterior computations
compared to Gibbs sampling and related MCMC methods, for any choice of smooth
link function.

\item \textbf{(Matrix-variate iid data). }The Bayesian exponential random
graph (ERG)\ model for network data is a doubly-intractable model. This model
is defined by a likelihood with an intractable normalizing constant
($\mathcal{Z}(\boldsymbol{\beta})$) formed by a sum over a large number of all
observable binary matrices of given dimension, and its posterior density
function is based on another intractable normalizing constant
\citep[e.g.,][]{CaimoFriel11}%
. Other doubly-intractable likelihood models include the Ising, Potts, Spatial
point process, Massive Gaussian Markov random field, and Mallows models. Due
to the intractability of $\mathcal{Z}(\boldsymbol{\beta})$, standard (e.g.,
MCMC)\ posterior estimation algorithms are inapplicable for doubly-intractable
models. Further, for such a model, it could be difficult to sample directly
from its likelihood function. Various MC posterior sampling methods have been
proposed for such models, which either eliminate or estimate $\mathcal{Z}%
(\boldsymbol{\beta})$ or $\mathcal{Z}(\boldsymbol{\beta})/\mathcal{Z}%
(\boldsymbol{\beta}^{\prime})$. But these methods are non-trivial to use and
not fully satisfactory
\citep{LiangEtAl16}%
. We will show that QIL\ can provide tractable posterior inferences for the
given doubly-intractable model, by specifying the QIL\ as a surrogate to the
model's implied low-dimensional logit model which does not depend on
$\mathcal{Z}(\boldsymbol{\beta})$. For example, the ERG model implies a
particular binary logit model for the given network data set
\citep{StraussIkeda90}%
. Also, we will show that a LASSO\ prior can lead to more reasonable marginal
posterior variance estimates for the ERG\ model parameters, compared to the
overly-high variance estimates that can be obtained from Maximum Likelihood
Estimation (MLE) methods\
\citep{CaimoFriel11}%
.

\item \textbf{(Multivariate iid data).} The Bayesian multivariate skew normal
distribution is an attractive model which is robust to empirical violations of
distributional symmetry.\ However, the model is defined by a likelihood
function that is rather unmanageable for parameter estimation purposes
\citep{LiseoParisi13,AzzaliniCapitanio99}%
. Also, in many statistical applications, including those where the ratio
($p/n$) of the number of variables to the sample size is large, it is of
interest to perform posterior distribution inferences of the inverse
covariance matrix $\boldsymbol{\Omega}=(\omega_{jk})_{p\times p}$
\citep[e.g.,][]{FanEtAl14}%
. This matrix describes the partial correlation $-\omega_{jk}/\sqrt
{\omega_{jj}\omega_{kk}}$ between $(Y_{j},Y_{k})$ after removing the linear
effect of the other $p-2$ variables, for all $1\leq j<k\leq p$; and describes
the partial variance $1/\omega_{jj}$ of $Y_{j}$ after removing the linear
effect of the other $p-1$ variables, for all $1\leq j\leq p$
\citep{Pourahmadi11}%
. A robust inference method for the inverse covariance matrix was only
developed from a frequentist (non-Bayesian)\ perspective
\citep{ZhaoLiu14}%
. We will show that after approximating the multivariate skew-normal
likelihood by the QIL, this model can provide posterior inferences of the
inverse-covariance matrix while ignoring nuisance parameters, through the use
of a computationally-fast and simple importance sampling algorithm which
avoids costly matrix inversions.

\item \textbf{(Multivariate non-iid data). }The Bayesian approach to the
multivariate Wallenius (noncentral hypergeometric) distribution
\citep{Wallenius63,Chesson76}
is useful for the analysis of individual choice data, where each person
chooses\ (without replacement)\ any number of objects from a total set of
objects (resp.) from mutually-exclusive categories
\citep{GrazianEtAl18}%
. But this model's exact likelihood contains a computationally-costly integral
for each person
\citep{GrazianEtAl18}%
. In this study, we will show that the QIL, as a surrogate to the Wallenius
model likelihood, can provide tractable and accurate posterior inferences for
the model's choice weight parameters. This is done by specifying this QIL\ to
depend on means and variances of the category counts given the Wallenius model
parameters. The original Bayesian Wallenius model
\citep{GrazianEtAl18}
assumes that the choice parameters are the same for all persons, implying that
the choice data observations are iid over persons. The current study
introduces a new hierarchical Bayesian Wallenius model which can handle
non-iid choice data by allowing for the choice weight parameters to vary
across persons. The QIL can easily be applied to provide tractable posterior
distribution inferences for this new model.
\end{enumerate}

\noindent Our QIL\ framework is suitable for Bayesian inference with the above
models, but extends beyond these applications and allows us to push further
the boundaries of the class of problems can be addressed by likelihood-free
methods. The software code that was used for all data analyses are available
from the first author as Supplementary Material.

\section{Quantile Implied Likelihood (QIL) for Bayesian Inference}

To start setting the notational scene, any Bayesian statistical model for a
set of $n$ iid data observations $\mathbf{y}_{i}$, given by $\mathcal{Y}%
_{n}=\{\mathbf{y}_{i}\}_{i=1}^{n}$, specifies a likelihood function
$f_{\boldsymbol{\theta}}(\mathcal{Y}_{n})=%
{\textstyle\prod\nolimits_{i=1}^{n}}
f_{\boldsymbol{\theta}}(\mathbf{y}_{i})$ (with $\dim(\mathbf{y})\geq1$, and
$\mathbf{y}=y$ if $\dim(\mathbf{y})=1$), with prior density function
$\pi(\boldsymbol{\theta})$ (cdf $\Pi(\boldsymbol{\theta})$) defined on the
parameter space $\Theta\subseteq%
\mathbb{R}
^{\dim(\boldsymbol{\theta})}$. The (exact)\ likelihood $f_{\boldsymbol{\theta
}}(\mathbf{y})$ has cdf $F_{\boldsymbol{\theta}}(\mathbf{y})$, and for
univariate $y$, quantile function $q_{\boldsymbol{\theta}}(\lambda
)=F_{\boldsymbol{\theta}}^{-1}(\lambda)$, for $\lambda\in\lbrack0,1]$.
According to Bayes' theorem, the data set $\mathcal{Y}_{n}$ updates the prior
to a posterior distribution, defined by the density function $\pi
(\boldsymbol{\theta\mid}\mathcal{Y}_{n})=f_{\boldsymbol{\theta}}%
(\mathcal{Y}_{n})\pi(\boldsymbol{\theta})/m(\mathcal{Y}_{n})$, with marginal
likelihood $m(\mathcal{Y}_{n})=%
{\textstyle\int}
f_{\boldsymbol{\theta}}(\mathcal{Y}_{n})\mathrm{d}\Pi(\boldsymbol{\theta})$,
and posterior predictive density function $f_{n}(\mathbf{y})=%
{\textstyle\int}
f_{\boldsymbol{\theta}}(\mathbf{y})\mathrm{d}\Pi(\boldsymbol{\theta\mid
}\mathcal{Y}_{n})$. These ideas naturally extend to a Bayesian model for
non-iid data, defined by likelihood $f_{\boldsymbol{\theta}}(\mathcal{Y}_{n})=%
{\textstyle\prod\nolimits_{k=1}^{K}}
{\textstyle\prod\nolimits_{i_{k}=1}^{n_{k}}}
f_{\boldsymbol{\theta},k}(\mathbf{y}_{i,k})$ for $K\geq1$ independent groups,
where $\mathbf{y}_{i,k}$ denotes the $i$th observation in the $k$th group.

The QIL, described next in $\S 2.1$-$\S 2.2$, provides approximation to a
model likelihood $f_{\boldsymbol{\theta}}$ that may be intractable. $\S 2.1$
defines and describes the QIL, and shows how this is constructed from the
quantiles of univariate iid or non-iid data, and from the corresponding
quantiles of the exact model likelihood. $\S 2.2$ describes how the univariate
QIL\ can be constructed from multivariate iid or non-iid data observations,
based on univariate quantiles of the Mahalanobis depths for these observations
(resp.). Throughout $\S 2$-$\S 3$ we mention concrete examples to further
illustrate the QIL.

\subsection{QIL for Univariate Data}

Consider a data set $\mathcal{Y}_{n}$ of size $n$, sampled as $\mathcal{Y}%
_{n}=\{y_{i}\}_{i=1}^{n}\overset{\text{iid}}{\sim}F$, where $F$ ($f$) is a
given but unknown true continuous cdf (pdf) on $\mathbb{R}$, with
corresponding quantile function $q(\lambda)=F^{-1}(\lambda)=\{y:F(y)=\lambda
\}$ defined for any cdf probability $0\leq\lambda\leq1$.

\textit{Example 1}. As an illustration, suppose that the data set is sampled
as $\mathcal{Y}_{n}=\{y_{i}\}_{i=1}^{n}\overset{\text{iid}}{\sim}F$, with the
true $F$ being the Student's $t$ distribution, having cdf $F(\cdot
)=F_{\boldsymbol{\theta}}(\cdot)$ defined by:%
\begin{equation}
F_{\boldsymbol{\theta}}(y)=\mathrm{T}(y\mid\mu,\sigma,\nu)=%
{\displaystyle\int\nolimits_{-\infty}^{y}}
\dfrac{\Gamma((\nu+1)/2)}{\sigma\sqrt{2\pi}\Gamma(\nu/2)}[(1/\nu)\{\nu
+(z-\mu)^{2}/\sigma^{2}\}]^{-(\nu+1)/2}\mathrm{d}z, \label{tcdf}%
\end{equation}
\newline with corresponding pdf likelihood:%
\begin{equation}
f_{\boldsymbol{\theta}}(y)=\mathrm{t}(y\mid\mu,\sigma,\nu)=\tfrac{\Gamma
((\nu+1)/2)}{\sigma\sqrt{2\pi}\Gamma(\nu/2)}[(1/\nu)\{\nu+(y-\mu)^{2}%
/\sigma^{2}\}]^{-(\nu+1)/2}, \label{tpdf}%
\end{equation}
and parameters $\boldsymbol{\theta}=(\mu,\sigma,4)$ of location ($\mu$), scale
($\sigma>0$), and degrees of freedom $\nu=2$. This Student's t distribution
$\mathrm{T}(\cdot\mid\mu,\sigma,2)$, for any cdf probability $\lambda\in
(0,1)$, has quantile function
\citep{Hill70}
given by:%
\begin{equation}
q(\lambda)=q_{\boldsymbol{\theta}}(\lambda)=F_{\boldsymbol{\theta}}%
^{-1}(\lambda)=\mathrm{T}^{-1}(\lambda\mid\mu,\sigma,2)=2(\lambda
-1/2)\sqrt{2/(4\lambda(1-\lambda))}\sigma+\mu. \label{qtcdf}%
\end{equation}
$\square\smallskip$

Now, for whatever true continuous distribution $F$ that happens to generate
the sample data, via $\mathcal{Y}_{n}\overset{\text{iid}}{\sim}F$ (not
necessarily a Student's distribution), the given sample data set
$\mathcal{Y}_{n}$ is fully described by the empirical cdf $\widehat{F}%
_{n}(\cdot)=\tfrac{1}{n}%
{\textstyle\sum\nolimits_{i=1}^{n}}
\mathbf{1}(y_{i}\leq\cdot)$ with $n$ order statistics given by
$\widehat{\mathbf{q}}_{n,n}=(\widehat{q}_{j})_{j=1}^{n}=(y_{(1)}%
<\cdots<y_{(n)})$.

The QIL is based on the subset of $d$ sample quantiles of data $\mathcal{Y}%
_{n}$. This enables a more efficient data analysis, especially when the sample
size $n$ is large. Specifically, the QIL\ is constructed from a subset of
$d\leq n$ quantiles $\widehat{\mathbf{q}}_{n,d}=(\widehat{q}_{j}%
=\widehat{q}(\lambda_{j}))_{j=1}^{d}$ of the sample data $\mathcal{Y}_{n}$, on
equally-spaced probabilities $\boldsymbol{\lambda}_{d}=(\tfrac{j}{n+1}%
)_{j=1}^{d}$ in $(0,1)$, which can be routinely found by linear interpolation
\citep[][Def. 5]{HyndmanFan96}%
. For large $n$, we can write $\widehat{\mathbf{q}}_{n,d}=(y_{(\lceil
\lambda_{j}n\rceil)})_{j=1}^{d}$, where $\lceil\cdot\rceil$ is the ceiling function.

The number of quantiles $d$ of the sample data $\mathcal{Y}_{n}$ can be
selected as the value $d(\epsilon;\widehat{F}_{n})$ which yields an empirical
distribution $\widehat{F}_{d}(y_{i})$ of the $d$ sample quantiles that
well-approximates the full data empirical distribution $\widehat{F}_{n}$, for
a chosen constant $\epsilon\geq0$ (e.g., $\epsilon=.01$). Specifically, the
value $d(\epsilon;\widehat{F}_{n})$ can be found by the minimizing solution:%
\begin{equation}
d(\epsilon;\widehat{F}_{n})=\min_{d\in\{1,\ldots,n\}}\{d:|\widehat{F}%
_{n}(y_{i})-\widehat{F}_{d}(y_{i})|\leq\epsilon,\;\epsilon\geq0\}.
\label{subSelect}%
\end{equation}

We now formally state the assumptions of the QIL\ for univariate data.\medskip

\textit{Assumption 1}. For the given (univariate)\ data set $\mathcal{Y}_{n}$,
its empirical distribution $\widehat{F}_{n}$ is well-approximated by the
distribution $\widehat{F}_{d}$ of $\widehat{\mathbf{q}}_{n,d}=(\widehat{q}%
(\lambda_{j}))_{j=1}^{d}$, the $d(\epsilon;\widehat{F}_{n})$ sample quantiles
of the data on equally-spaced (cdf) probabilities $\boldsymbol{\lambda}%
_{d}=(\tfrac{j}{n+1})_{j=1}^{d}$. $\square\smallskip$

\textit{Assumption 2}. For the given (univariate)\ data set $\mathcal{Y}_{n}$
generated by the unknown true distribution $F$, the specified Bayesian model
defined by likelihood cdf $F_{\boldsymbol{\theta}}$ (pdf
$f_{\boldsymbol{\theta}}$) is correct in the sense that the equality
$F=F_{\boldsymbol{\theta}}$ exists for some parameter $\boldsymbol{\theta}%
\in\Theta$ in the support of the model's prior $\pi(\boldsymbol{\theta})$.
$\square\smallskip$

The QIL is based on large-sample asymptotic theory. If Assumption 2 holds,
then for any $d\geq1$ and $\boldsymbol{\lambda}_{d}=(\lambda_{1}%
,\ldots,\lambda_{d})$ (with $0<\lambda<1$), the vector of sample quantiles
$\widehat{\mathbf{q}}_{n,d}=(y_{(\lceil\lambda_{j}n\rceil)})_{j=1}^{d}$ has a
$d$-variate normal distribution law ($\mathcal{L}$)
\citep[e.g.,][Ch.13]{Walker68,Ferguson96}%
, given by:%
\begin{equation}
n^{1/2}(\widehat{\mathbf{q}}_{n,d}-\mathbf{q}_{\boldsymbol{\theta}%
,d})\overset{\mathcal{L}}{\rightarrow}\mathrm{N}_{d}(\mathbf{0},\mathbf{V}%
_{f_{\boldsymbol{\theta}}})\text{ as }n\rightarrow\infty,
\end{equation}
with mean vector $\mathbf{q}_{\boldsymbol{\theta},d}=(q_{\boldsymbol{\theta}%
}(\lambda_{j})=F_{\boldsymbol{\theta}}^{-1}(\lambda_{j}))_{j=1}^{d}$, and
covariance matrix:%
\begin{equation}
\mathbf{V}(f_{\boldsymbol{\theta}})=\left(  \dfrac{\min(\lambda_{j}%
,\lambda_{k})[1-\max(\lambda_{j},\lambda_{k})]}{f_{\boldsymbol{\theta}%
}(q_{\boldsymbol{\theta}}(\lambda_{j}))f_{\boldsymbol{\theta}}%
(q_{\boldsymbol{\theta}}(\lambda_{k}))}\right)  _{d\times d}. \label{Covar}%
\end{equation}
Then, asymptotically, $\widehat{\mathbf{q}}_{n,d}\overset{\mathcal{L}%
}{\rightarrow}\mathrm{N}_{d}(\mathbf{q}_{\boldsymbol{\theta},d},\tfrac{1}%
{n}\mathbf{V}(f_{\boldsymbol{\theta}}))$ as $n\rightarrow\infty$, which
implies that the pivotal quantity:%
\begin{equation}
t_{\boldsymbol{\theta}}(\mathcal{Y}_{n})=n(\widehat{\mathbf{q}}_{n,d}%
-\mathbf{q}_{\boldsymbol{\theta},d})^{\intercal}[\mathbf{V}%
(f_{\boldsymbol{\theta}})]^{-1}(\widehat{\mathbf{q}}_{n,d}-\mathbf{q}%
_{\boldsymbol{\theta},d}) \label{Pivotal}%
\end{equation}
follows a chi-square ($\chi_{d}^{2}$) distribution on $d$ degrees of freedom,
that is, $t_{\boldsymbol{\theta}}(\mathcal{Y}_{n})\sim\chi_{d}^{2}$.

The statistic $t_{\boldsymbol{\theta}}(\mathcal{Y}_{n})$ in (\ref{Pivotal}) is
a pivotal quantity, because it is a function of the data observations
$\widehat{\mathbf{q}}_{n,d}$ and of the parameters $\boldsymbol{\theta}$, and
has the same ($\chi_{d}^{2}$) distribution for all parameters
$\boldsymbol{\theta}\in\Theta$
\citep[e.g.,][]{DeGrootSchervish12}%
. Also, when the Bayesian model likelihood $f_{\boldsymbol{\theta}}$ is
intractable, then $\mathbf{V}(f_{\boldsymbol{\theta}})$ in (\ref{Covar}) can
be calculated by $\mathbf{V}(f_{\boldsymbol{\theta}}^{\text{e}})$ using the
equiprobability pdf
\citep[][pp. 208-9]{Breiman73}%
:%
\begin{equation}
f_{\boldsymbol{\theta}}^{\text{e}}(\cdot)=%
{\displaystyle\sum\limits_{j=1}^{d}}
\dfrac{\mathbf{1}(q_{\boldsymbol{\theta}}(\lambda_{j-1})<\cdot\leq
q_{\boldsymbol{\theta}}(\lambda_{j}))}{(d+1)(q_{\boldsymbol{\theta}}%
(\lambda_{j})-q_{\boldsymbol{\theta}}(\lambda_{j-1}))}, \label{eqpDens}%
\end{equation}
where $\mathbf{1}(\cdot)$ is an the indicator function and $\lambda_{0}%
\equiv\epsilon>0$ is a small constant.\medskip

\textit{Example 1 (continued)}. Suppose again that data are generated as
$\mathcal{Y}_{n}=\{y_{i}\}_{i=1}^{n}\overset{\text{iid}}{\sim}F$, from the
Student's $t$ distribution with cdf $F(\cdot)=F_{\boldsymbol{\theta}}%
(\cdot)=\mathrm{T}(\cdot\mid\mu,\sigma,2)$. Then the sample quantiles
$\widehat{\mathbf{q}}_{n,d}=(y_{(\lceil\lambda_{j}n\rceil)})_{j=1}^{d}$ of the
data $\mathcal{Y}_{n}$, on $d$ equally-spaced cdf probabilities
$\boldsymbol{\lambda}_{d}=(\lambda_{j}=\tfrac{j}{n+1})_{j=1}^{d}$, have the
asymptotic $d$-variate normal distribution, $\widehat{\mathbf{q}}%
_{n,d}\overset{\mathcal{L}}{\rightarrow}\mathrm{N}_{d}(\mathbf{q}%
_{\boldsymbol{\theta},d},\tfrac{1}{n}\mathbf{V}(f_{\boldsymbol{\theta}}))$ as
$n\rightarrow\infty$. Here, the asymptotic mean quantile vector is given by
$\mathbf{q}_{\boldsymbol{\theta},d}=(q_{\boldsymbol{\theta}}(\lambda
_{1}),\ldots,q_{\boldsymbol{\theta}}(\lambda_{j})\ldots,q_{\boldsymbol{\theta
}}(\lambda_{d}))$, with quantile function $q_{\boldsymbol{\theta}}(\lambda)$
given by equation (\ref{qtcdf}), and the asymptotic covariance matrix
$\mathbf{V}(f_{\boldsymbol{\theta}})$ is given by (\ref{Covar}), based on
$\boldsymbol{\lambda}_{d}=(\lambda_{j}=\tfrac{j}{n+1})_{j=1}^{d}$ and on the
Student $\mathrm{T}(\cdot\mid\mu,\sigma,2)$ pdf given by (\ref{tpdf}).
$\square\smallskip$

For whatever true distribution $F$ that happens to generate the given data set
$\mathcal{Y}_{n}$, the $\chi_{d}^{2}$ distribution for the pivotal quantity
$t_{\boldsymbol{\theta}}$ (\ref{Pivotal}) is an asymptotic confidence
distribution, meaning that:%
\begin{equation}
\Pr[t_{\boldsymbol{\theta}}(\mathcal{Y}_{n})\leq\chi_{d}^{-2}(u)]=\Pr[\chi
_{d}^{2}(t_{\boldsymbol{\theta}}(\mathcal{Y}_{n}))\leq u]\overset{\mathcal{L}%
}{\rightarrow}u,\text{ as }n\rightarrow\infty, \label{ConfidenceDist}%
\end{equation}
holds for all $u\in(0,1)$ and all $(\mathcal{Y}_{n},\boldsymbol{\theta})$,
with cdf $\chi_{d}^{2}(t)$ and corresponding quantile function $\chi_{d}%
^{-2}(u)$, and stochastic confidence interval $(0,\chi_{d}^{-2}(u)]$ of
coverage probability $u=1-\alpha$
\citep{XieSingh13,NadarajahEtAl15}%
.

The QIL for univariate iid data is the $\chi_{d}^{2}$ pdf corresponding to the
confidence distribution cdf (\ref{ConfidenceDist}), and it is the asymptotic
implied likelihood confidence density
\citep{Efron93}
defined by:
\begin{equation}
f_{\boldsymbol{\theta}}^{\text{Q}}(\mathcal{Y}_{n})\equiv\frac{\mathrm{d}%
\chi_{d}^{2}(t_{\boldsymbol{\theta}}(\mathcal{Y}_{n}))}{\mathrm{d}%
t_{\boldsymbol{\theta}}(\mathcal{Y}_{n})}=\frac{[t_{\boldsymbol{\theta}%
}(\mathcal{Y}_{n})]^{d/2-1}}{\exp[%
\frac12
t_{\boldsymbol{\theta}}(\mathcal{Y}_{n})]2^{d/2}\Gamma(d/2)}, \label{QILiid}%
\end{equation}
and based on the pivotal function $t_{\boldsymbol{\theta}}(\mathcal{Y}_{n})$
of the model parameters $\boldsymbol{\theta}$ which depends on the total
sample size $n$ of the full data set $\mathcal{Y}_{n}$ (see (\ref{Pivotal})).
The QIL is not the `quantile likelihood'
\citep[][and references therein]{HeathcoteBrown04}
which contains computationally costly integrals.

The $\chi_{d}^{2}$ confidence distribution cdf (\ref{ConfidenceDist}) provides
an asymptotic\ distribution for the pivotal test statistic
$t_{\boldsymbol{\theta}}(\mathcal{Y}_{n})$ under a null hypothesis.
Specifically, this confidence distribution is for the test of the null
hypothesis $H_{0}:F=F_{\boldsymbol{\theta}}$ (or $\mathbf{q}_{d}%
=\mathbf{q}_{\boldsymbol{\theta},d}$)\ versus the alternative hypothesis
$H_{1}:F\neq F_{\boldsymbol{\theta}}$ (or $\mathbf{q}_{d}\neq\mathbf{q}%
_{\boldsymbol{\theta},d}$), with unknown $F$ (and $\mathbf{q}_{d}%
=(F^{-1}(\lambda_{j}))_{j=1}^{d}$) estimated by $\widehat{\mathbf{q}}_{n,d}$,
known $F_{\boldsymbol{\theta}}$ (and $\mathbf{q}_{\boldsymbol{\theta}%
,d}=(F_{\boldsymbol{\theta}}^{-1}(\lambda_{j}))_{j=1}^{d}$), and p-value
$1-\chi_{d}^{2}(t_{\boldsymbol{\theta}}(\mathcal{Y}_{n}))$. The QIL
$f_{\boldsymbol{\theta}}^{\text{Q}}(\mathcal{Y}_{n})$ in (\ref{QILiid}) is the
pdf of this $\chi_{d}^{2}$ null hypothesis distribution, which measures the
plausibility of any given parameter value $\boldsymbol{\theta}$ for the given
data set $\mathcal{Y}_{n}$, under the null $H_{0}:F=F_{\boldsymbol{\theta}}$.

For univariate non-iid data $\mathcal{Y}_{n}=\{\mathcal{Y}_{n_{k}}\}_{k=1}%
^{K}=$ $\{\{\mathbf{y}_{i,k}\}\}_{k=1}^{K}$ from $K\geq1$ independent groups,
with true distribution (exact likelihood pdf)\ given by:%
\begin{equation}
f_{\boldsymbol{\theta}}(\mathcal{Y}_{n})=%
{\displaystyle\prod\limits_{k=1}^{K}}
{\displaystyle\prod\limits_{i=1}^{n}}
f_{\boldsymbol{\theta},k}(\mathbf{y}_{i,k}),
\end{equation}
the QIL\ is defined by:%
\begin{equation}
f_{\boldsymbol{\theta}}^{\text{Q}}(\mathcal{Y}_{n})\equiv%
{\displaystyle\prod\limits_{k=1}^{K}}
f_{\boldsymbol{\theta},k}^{\text{Q}}(\mathcal{Y}_{n_{k}})=%
{\displaystyle\prod\limits_{k=1}^{K}}
\frac{\mathrm{d}\chi_{d_{k}}^{2}(t_{\boldsymbol{\theta}}(\mathcal{Y}_{n_{k}%
}))}{\mathrm{d}t_{\boldsymbol{\theta}}(\mathcal{Y}_{n_{k}})}=%
{\displaystyle\prod\limits_{k=1}^{K}}
\frac{[t_{\boldsymbol{\theta}}(\mathcal{Y}_{n_{k}})]^{d_{k}/2-1}}{\exp[%
\frac12
t_{\boldsymbol{\theta}}(\mathcal{Y}_{n_{k}})]2^{d/2}\Gamma(d/2)},
\label{QILnoniid}%
\end{equation}
with corresponding pivotal quantities:%
\begin{equation}
t_{\boldsymbol{\theta}}(\mathcal{Y}_{n_{k}})=n_{k}(\widehat{\mathbf{q}}%
_{n_{k},d}-\mathbf{q}_{\boldsymbol{\theta},d})^{\intercal}[\mathbf{V}%
(f_{\boldsymbol{\theta},k})]^{-1}(\widehat{\mathbf{q}}_{n_{k},d}%
-\mathbf{q}_{\boldsymbol{\theta},d}),\ k=1,\ldots,K, \label{noniidPivotal}%
\end{equation}
having independent chi-square distributions $t_{\boldsymbol{\theta}%
}(\mathcal{Y}_{n})\overset{\text{ind}}{\sim}\chi_{d_{k}}^{2}$, for
$k=1,\ldots,K$.

\subsection{QIL for Multivariate Data}

The QIL (\ref{QILiid}) or (\ref{QILnoniid}) can be extended to multivariate
iid or non-iid data. Consider a multivariate iid data set sampled as
$\mathcal{Y}_{n}=\{\mathbf{y}_{i}\}_{i=1}^{n}\overset{\text{iid}}{\sim}F$,
with unknown continuous cdf $F$ (pdf $f$) defined on $%
\mathbb{R}
^{p}$, $p=\dim(\mathbf{y})\geq1$. For a Bayesian model with likelihood
$f_{\boldsymbol{\theta}}$, the Mahalanobis depth function
\citep{LiuSingh93}
is defined by:%
\begin{equation}
D_{M}(\mathbf{y};\boldsymbol{\mu}_{\boldsymbol{\theta}},\mathbf{\Sigma
}_{\boldsymbol{\theta}})=[1+(\mathbf{y}-\boldsymbol{\mu}_{\boldsymbol{\theta}%
})^{\intercal}\mathbf{\Sigma}_{\boldsymbol{\theta}}^{-1}(\mathbf{y}%
-\boldsymbol{\mu}_{\boldsymbol{\theta}})]^{-1}=[1+M_{\boldsymbol{\theta}%
}(\mathbf{y})]^{-1},
\end{equation}
where $D_{M}:%
\mathbb{R}
^{p}\rightarrow%
\mathbb{R}
_{+}$, and $(\boldsymbol{\mu}_{\boldsymbol{\theta}},\mathbf{\Sigma
}_{\boldsymbol{\theta}})$ is the mean and covariance matrix of
$f_{\boldsymbol{\theta}}$ ($F_{\boldsymbol{\theta}}$), with
$M_{\boldsymbol{\theta}}(\mathbf{y})$ a Mahalanobis distance. The Mahalanobis
depth is consistent with four reasonable axioms for depth functions
\citep[see][]{Mosler13}%
. A key axiom is \textit{affine invariance}, which states that $D_{M}%
(\mathbf{A}\bullet+\,\mathbf{b};\boldsymbol{\mu}_{\boldsymbol{\theta}}^{\ast
},\mathbf{\Sigma}_{\boldsymbol{\theta}}^{\ast})=D_{M}(\bullet
;F_{\boldsymbol{\theta}})$ holds for any nonsingular matrix $\mathbf{A}$ and
constant vector $\mathbf{b}$, where $(\boldsymbol{\mu}_{\boldsymbol{\theta}%
}^{\ast},\mathbf{\Sigma}_{\boldsymbol{\theta}}^{\ast})$ is the mean and
covariance matrix of the distribution $F_{\boldsymbol{\theta}}^{\ast}$ of
$\mathbf{Ay+b}$ with $\mathbf{y}\sim F_{\boldsymbol{\theta}}$.

The Mahalanobis depth provides a coherent basis for multivariate quantiles,
and has the same probabilistic interpretations analogous to the univariate
case
\citep{Serfling02stN}%
. The rank order of the sample Mahalanobis depths $D_{M}(\mathbf{y}%
_{1};F_{\boldsymbol{\theta}}),\ldots,D_{M}(\mathbf{y}_{n}%
;F_{\boldsymbol{\theta}})$ define multivariate order statistics $\mathbf{y}%
_{(1)},\ldots,\mathbf{y}_{(n)}$
\citep[][p.787]{LiuEtAl99}%
. Here, $\mathbf{y}_{(1)}$ has the highest depth and defines a multivariate
median
\citep[][]{RousseeuwLeroy87}%
, and $\mathbf{y}_{(n)}$ has the smallest depth and is the most outlying
point. For the one-dimensional transformed data $\{\widehat{y}_{i}%
=D_{M}(\mathbf{y}_{i};\widehat{\boldsymbol{\mu}},\widehat{\mathbf{\Sigma}%
})\}_{i=1}^{n}$, it is possible to select $d(\epsilon)$ sample quantiles on
equally-spaced (cdf) probabilities $\boldsymbol{\lambda}_{d}=(\lambda
_{j})_{j=1}^{d}$ for some $\epsilon\geq0$ using (\ref{subSelect}) with
$\widehat{y}$ in place of $y$.\ Here, $(\widehat{\boldsymbol{\mu}%
},\widehat{\mathbf{\Sigma}})$ can be chosen as a robust estimator of the mean
and covariance matrix
\citep[e.g.,][]{RousseeuwVanDriessen99}%
.

The QIL\ for multivariate data has a tractable form, as a consequence of the
following assumption and theorem.\medskip

\textit{Assumption 3.} For the given multivariate\ data set $\mathcal{Y}_{n}$
sampled from the unknown true distribution $F$, the specified Bayesian model
defined by likelihood $F_{\boldsymbol{\theta}}$ (with mean and covariance
$(\boldsymbol{\mu}_{\boldsymbol{\theta}},\mathbf{\Sigma}_{\boldsymbol{\theta}%
})$) is correct in the sense that $F=F_{\boldsymbol{\theta}}$ for some
parameter $\boldsymbol{\theta}\in\Theta$ in the support of the prior
$\pi(\boldsymbol{\theta})$, and $F_{\boldsymbol{\theta}}$ for any
$\boldsymbol{\theta}\in\Theta$ can be well-approximated by a multivariate skew
normal distribution $\mathrm{SN}_{p}(\boldsymbol{\xi},\boldsymbol{\Sigma
},\boldsymbol{\alpha})$, defined by the pdf:%
\begin{equation}
\mathrm{sn}_{p}(\mathbf{y};\boldsymbol{\xi},\boldsymbol{\Sigma}%
,\boldsymbol{\alpha})=2\phi_{p}(\mathbf{y}-\boldsymbol{\xi};\boldsymbol{\Sigma
})\Phi(\boldsymbol{\alpha}^{\intercal}\boldsymbol{\omega}^{-1}(\mathbf{y}%
-\boldsymbol{\xi})),\newline\label{SNpdf}%
\end{equation}
with $\phi_{p}$ a zero-mean $p$-variate normal pdf, \textrm{Normal}$(0,1)$ cdf
$\Phi$, and parameters $(\boldsymbol{\xi},\boldsymbol{\alpha}%
,\boldsymbol{\Sigma})$ of location $\boldsymbol{\xi}\in\mathbb{R}^{p}$, shape
$\boldsymbol{\alpha}\in\mathbb{R}^{p}$, and normal $p\times p$ covariance
matrix $\boldsymbol{\Sigma}=\boldsymbol{\omega\Sigma}_{z}\boldsymbol{\omega}$
\citep{AzzaliniCapitanio99}%
, where $\mathbf{\Sigma}_{\boldsymbol{\theta}}=\boldsymbol{\Sigma}$.
$\square\smallskip$

\begin{thm}
If Assumption 3 holds and $\boldsymbol{Y}\sim F_{\boldsymbol{\theta}}$,
then:\newline(a) $\ M_{\boldsymbol{\theta}}(\boldsymbol{Y})=(\boldsymbol{Y}%
-\boldsymbol{\mu}_{\boldsymbol{\theta}})^{\intercal}\boldsymbol{\Sigma
}_{\boldsymbol{\theta}}^{-1}(\boldsymbol{Y}-\boldsymbol{\mu}%
_{\boldsymbol{\theta}})\sim\chi_{p}^{2}$;\newline(b) \ the Mahalanobis depth,
for a fixed $\mathbf{y}$, has cdf and monotone transformation $D_{R}$ given by
the complementary $\chi_{p}^{2}$ cdf:
\begin{subequations}
\label{DR}%
\begin{align}
D_{R}(\mathbf{y};\boldsymbol{\mu}_{\boldsymbol{\theta}},\mathbf{\Sigma})  &
=\text{Pr}_{F_{\boldsymbol{\theta}}}(D_{M}(\boldsymbol{Y};\boldsymbol{\mu
}_{\boldsymbol{\theta}},\mathbf{\Sigma}_{\boldsymbol{\theta}})\leq
D_{M}(\mathbf{y};\boldsymbol{\mu}_{\boldsymbol{\theta}},\mathbf{\Sigma
}_{\boldsymbol{\theta}}))\label{DR1}\\
&  =\text{Pr}_{F_{\boldsymbol{\theta}}}(M_{\boldsymbol{\theta}}(\boldsymbol{Y}%
)\geq M_{\boldsymbol{\theta}}(\mathbf{y}))=1-\chi_{p}^{2}%
(M_{\boldsymbol{\theta}}(\mathbf{y})); \label{DR2}%
\end{align}
(c)\ \ $D_{R}(\boldsymbol{Y};\boldsymbol{\mu}_{\boldsymbol{\theta}%
},\mathbf{\Sigma}_{\boldsymbol{\theta}})\sim$ \textrm{Uniform}$(0,1)$.
\end{subequations}
\end{thm}

\textit{Proof}. Outcome (a)\ holds since $\boldsymbol{Y}\sim\mathrm{SN}%
_{p}(\boldsymbol{\xi},\boldsymbol{\Sigma},\boldsymbol{\alpha})$ (with
$\boldsymbol{\Sigma=\Sigma}_{\boldsymbol{\theta}}$) implies that
$(\boldsymbol{Y}-\boldsymbol{\mu}_{\boldsymbol{\theta}})\sim\mathrm{SN}%
_{p}(\mathbf{y};\mathbf{0},\boldsymbol{\Sigma},\boldsymbol{\alpha})$, and that
by Proposition 5 of
\citet{AzzaliniCapitanio99}%
, $\boldsymbol{Z}\sim\mathrm{SN}_{p}(\mathbf{y};\mathbf{0},\boldsymbol{\Sigma
}_{\boldsymbol{Z}},\boldsymbol{\alpha}_{\boldsymbol{Z}})$ with $\boldsymbol{Z}%
=\boldsymbol{\Sigma}^{-1}(\boldsymbol{Y}-\boldsymbol{\mu}_{\boldsymbol{\theta
}})$ and $\boldsymbol{\Sigma}_{\boldsymbol{Z}}=\boldsymbol{\Sigma}%
^{-1/2}\boldsymbol{\Sigma\Sigma}^{-1/2}=\mathbf{I}_{p}$, leading to $Z_{k}%
\sim\mathrm{SN}_{p}(0,1,\alpha_{Z_{k}})$ and $Z_{k}^{2}\sim\chi_{1}^{2}$
independently for $k=1,\ldots,p$, with $\boldsymbol{Z}^{\intercal
}\boldsymbol{Z}=M_{\boldsymbol{\theta}}(\boldsymbol{Y})$. Outcome (b)\ holds
true, because as a consequence of the skew-normality assumption for
$F_{\boldsymbol{\theta}}$ and the affine invariance and strict monotonicity
properties of the Mahalanobis depth $D_{M}$ (implying the affine invariance of
$D_{R}$), the contours of constant $D$ are of the form $(\mathbf{y}%
-\boldsymbol{\mu}_{\boldsymbol{\theta}})^{\intercal}\mathbf{\Sigma}%
^{-1}(\mathbf{y}-\boldsymbol{\mu}_{\boldsymbol{\theta}})=c$
\citep[][Prop. 3.1]{LiuSingh93}%
. It is easy to verify that for the family of skew-normal distributions, the
depth $D_{M}(\mathbf{y};\boldsymbol{\mu}_{\boldsymbol{\theta}},\mathbf{\Sigma
})$ satisfies \textit{strict monotonicity} in the sense that there is no
neighborhood of $\mathbf{y}$ for which the depth takes on a constant value
\citep[][Def.3.1]{Dyckerhoff17}%
. (c)\ follows from the probability integral transform, given (b)\ which
establishes the continuity of the Mahalanobis depth distribution
\citep[][Theorem 5.2]{LiuSingh93}%
. $\square\smallskip$

If Assumption 3 holds, then the sample quantiles $\widehat{\mathbf{q}}%
_{n,d}=(D_{R}(\mathbf{y}_{j};\boldsymbol{\mu}_{\boldsymbol{\theta}%
},\mathbf{\Sigma}_{\boldsymbol{\theta}}))_{j=1}^{d}$ on $d$ cdf probabilities
$\boldsymbol{\lambda}_{d}=(\lambda_{j})_{j=1}^{d}$ have an asymptotic normal
distribution
\citep[][\S4.3]{Serfling02jma}%
, here, with mean $\mathbf{q}_{\boldsymbol{\theta},d}=(F_{R,\boldsymbol{\theta
}}^{-1}(\lambda_{j}))_{j=1}^{d}=\boldsymbol{\lambda}_{d}$ and covariance
matrix:
\begin{equation}
\mathbf{V}(f_{\boldsymbol{\theta}})=%
\genfrac{(}{)}{}{0}{\min(\lambda_{j},\lambda_{k})[1-\max(\lambda_{j}%
,\lambda_{k})]}{f_{R,\boldsymbol{\theta}}(q_{R,\boldsymbol{\theta}}%
(\lambda_{j}))f_{R,\boldsymbol{\theta}}(q_{R,\boldsymbol{\theta}}(\lambda
_{k}))}%
_{d\times d}=\left(  \min(\lambda_{j},\lambda_{k})(1-\max(\lambda_{j}%
,\lambda_{k}))\right)  _{d\times d},
\end{equation}
with $f_{R,\boldsymbol{\theta}}(r)=\mathbf{1}(0<r<1)$ and
$q_{R,\boldsymbol{\theta}}(\lambda)=F_{R,\boldsymbol{\theta}}^{-1}%
(\lambda)=\lambda$. That is, under Assumption 3, the QIL\ for multivariate
data can be simply constructed from the pdf and quantile function of the
\textrm{Uniform}$(0,1)$ distribution.

It then follows that for multivariate iid observations, the QIL is still given
by (\ref{QILiid}), with $\chi_{d}^{2}$ distributed pivotal statistic
(\ref{Pivotal}). Then the direct connections between the confidence
distribution and hypothesis testing of $H_{0}:F=F_{\boldsymbol{\theta}}$ (vs.
$H_{1}:F\neq F_{\boldsymbol{\theta}}$) still remain. For the multivariate
setting, such a test is sensitive to location or dispersion departures from
$F_{\boldsymbol{\theta}}$
\citep{LiuSingh93}%
. Also, by extension, for multivariate non-iid observations in $K$\ groups,
the QIL\ is still given by the likelihood (\ref{QILnoniid}) with corresponding
pivotal quantities (\ref{noniidPivotal}).

Assumption 3 of QIL\ may seem potentially overly restrictive because it
suggests the assumption that the true underlying data generating distribution
$F$ is from a multivariate skew normal family. This assumption happens to be
reasonable for the multivariate models described in $\S 3.5$-$\S 3.6$.
However, this assumption does not preclude the possibility that true
distribution $F$ is from the space of continuous multivariate distributions,
which can be supported by a mixture model with parameters $\boldsymbol{\theta
}=(\boldsymbol{\theta}^{\ast},(d_{i})_{i=1}^{n})$ that include latent mixture
component (cluster) membership variables $d_{i}$ for observations indexed by
$i=1,...,n$ (resp.). Then, conditionally on these latent variables, the
multivariate likelihood densities $f_{\boldsymbol{\theta}^{\ast},d_{i}}$, for
$i=1,...,n$ (with $i=k$) can be reasonably be assumed to be in the support of
multivariate skew normal distribution. This is true according to common
Bayesian mixture models such as mixtures of multivariate normal distributions.
This means that the QIL for multivariate non-iid observations can be
applicable to Bayesian mixture models. These ideas are outside the scope of
this paper but worthy of consideration in future work.

\subsection{The Posterior Distribution Based on the QIL}

For any Bayesian model, the QIL\ $f_{\boldsymbol{\theta}}^{\text{Q}}$, a
surrogate for the exact model likelihood, combines with the model's prior
distribution $\pi(\boldsymbol{\theta})$ to yield an approximate\ posterior
distribution $\pi_{\text{Q}}(\boldsymbol{\theta\mid}\mathcal{Y}_{n})\propto
f_{\boldsymbol{\theta}}^{\text{Q}}(\mathcal{Y}_{n})\pi(\boldsymbol{\theta})$
(cdf $\Pi_{\text{Q}}(\boldsymbol{\theta\mid}\mathcal{Y}_{n})$) and predictive
density $f_{n}^{\text{Q}}(y)=%
{\textstyle\int}
f_{\boldsymbol{\theta}}^{\text{Q}}(\mathcal{Y}_{n})\mathrm{d}\Pi_{\text{Q}%
}(\boldsymbol{\theta\mid}\mathcal{Y}_{n})$. From the approximate posterior
distribution $\pi_{\text{Q}}(\boldsymbol{\theta\mid}\mathcal{Y}_{n})$, the
maximum a posteriori (MAP) estimate and corresponding posterior covariance
matrix of $\boldsymbol{\theta}$ can be estimated by minimizing a negative
penalized log-likelihood. Also, any function of the full posterior
distribution $\pi_{\text{Q}}(\boldsymbol{\theta\mid}\mathcal{Y}_{n})$ can be
estimated using any appropriate standard Monte Carlo algorithm. For the
examples of models covered in $\S 3$, we primarily consider an adaptive
Metropolis-Hastings algorithm, and an importance sampling algorithm, which are
further described in $\S 2.3.1$-$\S 2.3.2$.

As an aside, under the uninformative flat prior $\pi(\boldsymbol{\theta
})\propto1$ for the model parameters $\boldsymbol{\theta}$, the posterior
distribution $\pi_{\text{Q}}(t_{\boldsymbol{\theta}}\mid\mathcal{Y}_{n})$
arising from $\pi_{\text{Q}}(\boldsymbol{\theta\mid}\mathcal{Y}_{n})$ is based
on a matching prior distribution
\citep{WelchPeers63}
which coincides with the confidence interval system $\chi_{d}^{-2}(u)$ of the
pivotal quantity $t_{\boldsymbol{\theta}}$. Then by definition, for all
$0<u<1$, the QIL\ $f_{\boldsymbol{\theta}}^{\text{Q}}(\mathcal{Y}%
_{n};t_{\boldsymbol{\theta}})=\frac{\mathrm{d}\chi_{d}^{2}%
(t_{\boldsymbol{\theta}}(\mathcal{Y}_{n}))}{\mathrm{d}t_{\boldsymbol{\theta}%
}(\mathcal{Y}_{n})}$ and the posterior density $\pi(t_{\boldsymbol{\theta}%
}\mid\mathcal{Y}_{n})$ must satisfy
\citep[][p.201]{EfronHastie16}%
:%
\begin{equation}%
{\displaystyle\int\nolimits_{0}^{\chi_{d}^{-2}(u)}}
\frac{\mathrm{d}\chi_{d}^{2}(t_{\boldsymbol{\theta}}(\mathcal{Y}_{n}%
))}{\mathrm{d}t_{\boldsymbol{\theta}}(\mathcal{Y}_{n})}\mathrm{d}%
t_{\boldsymbol{\theta}}=%
{\displaystyle\int\nolimits_{0}^{\chi_{d}^{-2}(u)}}
\pi_{\text{Q}}(t_{\boldsymbol{\theta}}\mid\mathcal{Y}_{n})\mathrm{d}%
t_{\boldsymbol{\theta}}=u. \label{postEqual}%
\end{equation}
This relation (\ref{postEqual}) holds for all $\boldsymbol{\theta}\in\Theta$,
because $t_{\boldsymbol{\theta}}$ is a pivotal quantity which by definition
has the same ($\chi_{d}^{2}$) distribution for all $\boldsymbol{\theta}$.

\subsubsection{MAP and\ Posterior Covariance Estimation Using Penalized QIL}

The mode of the posterior distribution $\pi_{\text{Q}}(\boldsymbol{\theta\mid
}\mathcal{Y}_{n})$ defines the QIL\ MAP estimator $\widehat{\boldsymbol{\theta
}}_{\text{Q}}$, which coincides with the mean when the posterior is unimodal
and symmetric. The posterior covariance is the Hessian matrix inverse
evaluated at the mode. Under the flat prior $\pi(\boldsymbol{\theta})\propto
1$, the QIL\ MAP estimator $\widehat{\boldsymbol{\theta}}_{\text{Q}}$
approximates the MLE\ ($\widehat{\boldsymbol{\theta}}_{\text{MLE}}=\arg
\max_{\boldsymbol{\theta}\in\Theta}$ $f_{\boldsymbol{\theta}}(\mathcal{Y}%
_{n})$) having standard errors (SEs) given by the square-roots of the diagonal
elements of the Fisher information matrix inverse. As the sample size grows
$n\rightarrow\infty$, the posterior $\pi_{\text{Q}}(\boldsymbol{\theta\mid
}\mathcal{Y}_{n})$ approaches a multivariate normal distribution with mean
$\widehat{\boldsymbol{\theta}}_{\text{Q}}$ and covariance equal to $1/n$ times
the Fisher information matrix inverse
\citep[e.g.,][Ch.21]{Ferguson96}%
.

For any Bayesian model, the MAP estimate $\widehat{\boldsymbol{\theta}%
}_{\text{Q}}$ is the solution which maximizes the penalized log-likelihood,
which can be obtained by the minimizing solution:%
\begin{equation}
\widehat{\boldsymbol{\theta}}_{\text{Q}}(\mathcal{Y}_{n}%
)=\,\underset{\boldsymbol{\theta}\in\Theta}{\arg\min}\left[  -%
{\displaystyle\sum\limits_{k=1}^{K}}
\log\frac{[t_{\boldsymbol{\theta}}(\mathcal{Y}_{n_{k}})]^{d/2-1}}{\exp[%
\frac12
t_{\boldsymbol{\theta}}(\mathcal{Y}_{n_{k}})]}-\log\pi(\boldsymbol{\theta
})\right]  . \label{QLEpen}%
\end{equation}
If the degrees of freedom $d$ takes on values of 1 or 2, then the QIL\ MAP
solution $\widehat{\boldsymbol{\theta}}_{\text{Q}}$ (\ref{QLEpen}) resembles a
minimum chi-square estimator
\citep[][p.152]{Ferguson96}
because then the $\chi_{d}^{2}$ pdf defining the QIL\ is skewed with mode $0$.
If $d\geq5$ and $T\sim\chi_{d}^{2}$, then approximately $(T/d)^{1/3}%
\sim\mathrm{N}(1-\tfrac{2}{9d},\tfrac{2}{9d})$
\citep{WilsonHilferty31}%
. Then the QIL\ MAP\ estimate $\widehat{\boldsymbol{\theta}}_{\text{Q}}$ can
be obtained as the penalized least-squares solution:%
\begin{equation}
\widehat{\boldsymbol{\theta}}_{\text{Q}}(\mathcal{Y}_{n}%
)=\,\underset{\boldsymbol{\theta}\in\Theta}{\arg\min}\left[
{\displaystyle\sum\limits_{k=1}^{K}}
\left\{  \left(  \frac{t_{\boldsymbol{\theta}}(\mathcal{Y}_{n_{k}})}{d_{k}%
}\right)  ^{1/3}-\left(  1-\frac{2}{9d_{k}}\right)  \right\}  ^{2}-\log
\pi(\boldsymbol{\theta})\right]  . \label{QLEpen2}%
\end{equation}
A suitable algorithm can be used to quickly compute the MAP\ estimate via the
solution (\ref{QLEpen}) or (\ref{QLEpen2}), and the corresponding Hessian
matrix, when the algorithm is initiated with a good starting value of
$\boldsymbol{\theta}$ (e.g., $\widehat{\boldsymbol{\theta}}_{\text{MLE}}$).\ A
good starting value can be easily set when the posterior $\pi_{\text{Q}%
}(\boldsymbol{\theta\mid}\mathcal{Y}_{n})$ is unimodal or low-dimensional.
Otherwise, multiple runs of this algorithm are needed for several plausible
starting values of $\boldsymbol{\theta}$ (resp.), in order to obtain
$\widehat{\boldsymbol{\theta}}_{\text{Q}}$ as the optimal minimizing solution
over the multiple runs. This approach has some risk of finding a local
minimum, depending on how well the starting values are chosen.

\subsubsection{Monte Carlo (MC) Algorithms for QIL-based Posterior
Distribution Inference}

Two alternative and standard Monte Carlo (MC)\ algorithms include an adaptive
random-walk Metropolis-Hastings (AM)\ algorithm
\citep[][\S2]{RobertsRosenthal09}%
, and the Vanilla Importance Sampling (VIS)\ algorithm
\citep{MengersenPudloRobert13}%
. Each algorithm, for any parameter function $h(\boldsymbol{\theta})$ of
interest, can be employed to produce a\ sample $\{h(\boldsymbol{\theta}%
_{s})\}_{s=1}^{S}$ with an average $\overline{h}_{S}$ that converges to the
target posterior expectation $\mathbb{E}_{\pi_{\text{Q}}(\boldsymbol{\theta
\mid}\mathcal{Y}_{n})}[h(\boldsymbol{\theta})]=%
{\textstyle\int}
h(\boldsymbol{\theta})\mathrm{d}\Pi_{\text{Q}}(\boldsymbol{\theta\mid
}\mathcal{Y}_{n})$ as $S\rightarrow\infty$. These algorithms are briefly
reviewed as follows, with more details provided by the cited references.

The AM algorithm, at each sampling iteration $s=1,2,\ldots,S$ of a $S$
algorithm run, generates a proposal $\boldsymbol{\theta}_{\ast}$ from the
multivariate normal mixture proposal distribution:
\begin{equation}
\boldsymbol{\theta}_{\ast}\sim w_{s}\mathrm{N}_{q}(\boldsymbol{\theta}%
_{s-1},\tfrac{2.38^{2}}{\dim(\boldsymbol{\theta})}\widehat{\Sigma}%
_{s})+(1-w_{s})\mathrm{N}_{q}(\boldsymbol{\theta}_{s-1},\tfrac{.01^{2}}%
{\dim(\boldsymbol{\theta})}\mathbf{I}_{q}),
\end{equation}
and then accepts it with $\boldsymbol{\theta}_{s}=\boldsymbol{\theta}_{\ast}$
with probability $\min\{1,\frac{f_{\boldsymbol{\theta}_{\ast}}^{\text{Q}%
}(\mathcal{Y}_{n})\pi(\boldsymbol{\theta}_{\ast})}{f_{\boldsymbol{\theta
}_{s-1}}^{\text{Q}}(\mathcal{Y}_{n})\pi(\boldsymbol{\theta}_{s-1})}\},$ and
otherwise rejects the proposal with $\boldsymbol{\theta}_{s}%
=\boldsymbol{\theta}_{s-1}$. Here, $w_{s}=.95\cdot\mathbf{1}(s>2\dim
(\boldsymbol{\theta}))$, $\widehat{\Sigma}_{s}$ is the covariance matrix of
the previously accepted samples $\{\boldsymbol{\theta}_{t}\}_{t=1}^{s-1}$, and
$q=\dim(\boldsymbol{\theta})$. The proposal distribution $\mathrm{N}%
_{q}(\boldsymbol{\theta}_{s-1},\tfrac{2.38^{2}}{\dim(\boldsymbol{\theta}%
)}\widehat{\Sigma}_{s})$ is used to approximate $\mathrm{N}_{q}%
(\boldsymbol{\theta}_{s-1},\tfrac{2.38^{2}}{\dim(\boldsymbol{\theta})}\Sigma)$
which is known to be optimal in a particular high dimensional context
\citep{RobertsRosenthal01}%
. Convergence can be accelerated by setting the algorithm's starting value
$\boldsymbol{\theta}_{0}$ equal to the MLE $\widehat{\boldsymbol{\theta}%
}_{\text{MLE}}$ or QIL\ MAP estimate $\widehat{\boldsymbol{\theta}}_{\text{Q}%
}$. This algorithm is most suitable for a posterior distribution that has at
least roughly-elliptical contours, as when the posterior approaches normality
when the sample size $n$ is large.

The VIS algorithm generates iid prior samples $\{\boldsymbol{\theta}%
_{s}\}_{s=1}^{S}\overset{\text{iid}}{\sim}\pi(\boldsymbol{\theta})$ and then
respectively sets their weights by $\{\omega_{s}=f_{\boldsymbol{\theta}_{s}%
}^{\text{Q}}(\mathcal{Y}_{n})\}_{s=1}^{S}$. This algorithm can be easily
parallelized to increase savings in computational time. From the VIS output,
the weighted estimator $\overline{h}_{S}=\left.
{\textstyle\sum\nolimits_{s=1}^{S}}
h(\boldsymbol{\theta}_{s})\overline{\omega}_{s}\right/
{\textstyle\sum\nolimits_{s=1}^{S}}
\overline{\omega}_{s}$ has finite variance and converges $\overline{h}%
_{S}\overset{a.s.}{\rightarrow}\mathbb{E}_{\pi_{\text{Q}}(\boldsymbol{\theta
\mid}\mathcal{Y}_{n})}[h(\boldsymbol{\theta})]$ by the strong law of large
numbers, where $\overline{\omega}_{s}=\omega_{s}/%
{\textstyle\sum\nolimits_{s=1}^{S}}
\omega_{s}$, because the prior $\pi(\boldsymbol{\theta})$ (instrumental
density) has thicker tails than those of $\pi_{\text{Q}}(\boldsymbol{\theta
\mid}\mathcal{Y}_{n})$, and $\sup$p$(\pi_{\text{Q}}(\boldsymbol{\theta\mid
}\mathcal{Y}_{n}))\subset\sup$p$(g(\boldsymbol{\theta}))$
\citep[][\S3.3.2]{CasellaRobert98,RobertCasella04}%
. The convergence of IS output can be evaluated by the Effective Sample Size
(ESS) statistic, $\mathrm{ESS}=1/%
{\textstyle\sum\nolimits_{s=1}^{S}}
\overline{\omega}_{s}^{2}$, $1\leq\mathrm{ESS}\leq S$. Here, $\mathrm{ESS}=1$
refers to a poor outcome, and $\mathrm{ESS}=S$ refers to the perfect outcome
indicating iid posterior samples $\{\boldsymbol{\theta}_{s}\}_{s=1}%
^{S}\overset{\text{iid}}{\sim}\pi_{\text{Q}}(\boldsymbol{\theta\mid
}\mathcal{Y}_{n})$
\citep{Liu01}%
.

\section{Illustrations of QIL}

We now illustrate the QIL\ through the analysis of many simulated data sets
and real data sets, using $27$ Bayesian models mentioned in $\S 1$. The
illustrations will involve comparisons with other model likelihoods (ABC, SL,
exact) and various algorithms for estimating from the posterior distribution.
Over all these data sets, the sample size ranged from $56$ to $10\,000\,000$,
and the number of covariates ranged from $0$ to $100$. All reported
computation times were obtained from a Intel i7 2.8GHz 16 GB\ RAM computer.
For all analyses based on MCMC\ sampling algorithms, convergence of the
generated samples to the posterior distribution was diagnosed according to
univariate trace plots, which displayed good mixing of each model parameter
over the sampling iterations.

\subsection{Standard Models for Univariate iid Data}

In order to provide a basic benchmark study of QIL, we performed a simulation
study involving 19 familiar probability distribution models (cdfs) for
univariate iid data
\citep[e.g., see][]
{JohnsonEtAl95v1and2,JohnsonKempKotz05}%
. For each model, Bayesian data analysis was performed using a rather
uninformative prior $\pi(\boldsymbol{\theta})$ for the parameters
$\boldsymbol{\theta}$. Table 1 lists these 19 models and corresponding priors.

\begin{center}%
\footnotesize
\begin{tabular}
[c]{llcllc}%
\multicolumn{6}{l}{}\\
\multicolumn{6}{l}{\textbf{Table 1.} Model, prior, and true data generating
parameters used for the data simulations.}\\\hline\hline
\multicolumn{1}{|l}{Model} & \multicolumn{1}{|l}{Prior, $\pi
(\boldsymbol{\theta})\propto$} & \multicolumn{1}{|c}{true $\boldsymbol{\theta
}$} & \multicolumn{1}{|l}{Model} & \multicolumn{1}{|l}{Prior, $\pi
(\boldsymbol{\theta})\propto$} & \multicolumn{1}{|c|}{true $\boldsymbol{\theta
}$}\\\hline
\multicolumn{1}{|l}{\textrm{Bernoulli}$(\theta)$} &
\multicolumn{1}{|c}{$\mathbf{1}(0<\theta<1)$} & \multicolumn{1}{|c}{$1/3$} &
\multicolumn{1}{|l}{\textrm{LN}$(\mu,\sigma)$} & \multicolumn{1}{|c}{%
\begin{tabular}
[c]{c}%
$\mathbf{1}(-10\leq\mu\leq10)$\\
$\times\mathbf{1}(0\leq\sigma\leq10^{2})$%
\end{tabular}
} & \multicolumn{1}{|c|}{$(3,1)$}\\\hline
\multicolumn{1}{|l}{\textrm{Beta}$(\alpha,\beta)$} &
\multicolumn{1}{|c}{$\mathbf{1}(0\leq\alpha,\beta\leq10^{2})$} &
\multicolumn{1}{|c}{$(3,1)$} & \multicolumn{1}{|l}{\textrm{NB}$(\theta,r)$} &
\multicolumn{1}{|c}{$\mathbf{1}(0<\theta<1)$} & \multicolumn{1}{|c|}{$(1/3,3)$%
}\\\hline
\multicolumn{1}{|l}{\textrm{BS}$(\beta,\gamma)$} &
\multicolumn{1}{|c}{$\mathbf{1}(0\leq\beta,\gamma\leq10^{2})$} &
\multicolumn{1}{|c}{$(3,1)$} & \multicolumn{1}{|l}{\textrm{N}$(\mu,1)$} &
\multicolumn{1}{|c}{$\exp[-\mu^{2}/(2(10^{2}))]$} &
\multicolumn{1}{|c|}{$(3,1)$}\\\hline
\multicolumn{1}{|l}{\textrm{Burr}$(\alpha,\varsigma,\kappa)$} &
\multicolumn{1}{|c}{$\mathbf{1}(0\leq\alpha,\varsigma,\kappa\leq10^{2})$} &
\multicolumn{1}{|c}{$(1/2,2,5)$} & \multicolumn{1}{|l}{\textrm{N}%
$(3,\sigma^{2})$} & \multicolumn{1}{|c}{$e^{-1/\sigma^{2}}\exp(-\sigma^{-2})$}
& \multicolumn{1}{|c|}{$(3,1)$}\\\hline
\multicolumn{1}{|l}{\textrm{Exp}$(\theta)$} & \multicolumn{1}{|c}{$\exp
(-\theta)$} & \multicolumn{1}{|c}{$3$} & \multicolumn{1}{|l}{\textrm{N}%
$(\mu,\sigma^{2})$} & \multicolumn{1}{|c}{$\exp[-\mu^{2}/(2\sigma^{2}%
10^{2})-\sigma^{-2}]$} & \multicolumn{1}{|c|}{$(3,1)$}\\\hline
\multicolumn{1}{|l}{\textrm{Ga}$(\alpha,\beta)$} &
\multicolumn{1}{|c}{$\mathbf{1}(0\leq\alpha,\beta\leq10^{2})$} &
\multicolumn{1}{|c}{$(3,1)$} & \multicolumn{1}{|l}{\textrm{Poi}$(\theta)$} &
\multicolumn{1}{|c}{$\exp(-\theta)$} & \multicolumn{1}{|c|}{$3$}\\\hline
\multicolumn{1}{|l}{\textrm{Geom}$(\theta)$} & \multicolumn{1}{|c}{$\mathbf{1}%
(0<\theta<1)$} & \multicolumn{1}{|c}{$1/3$} & \multicolumn{1}{|l}{\textrm{T}%
$(\mu,\sigma,\nu)$} & \multicolumn{1}{|c}{%
\begin{tabular}
[c]{c}%
$\exp(-\mu^{2}/(2(10^{2}))-\sigma)$\\
$\times\mathbf{1}(3\leq\nu\leq40)$%
\end{tabular}
} & \multicolumn{1}{|c|}{$(3,1,4)$}\\\hline
\multicolumn{1}{|l}{\textrm{GEV}$(\kappa,\sigma,\mu)$} & \multicolumn{1}{|c}{%
\begin{tabular}
[c]{c}%
$\mathbf{1}(-10\leq\kappa,\mu\leq10)$\\
$\times\mathbf{1}(\sigma>0)$%
\end{tabular}
} & \multicolumn{1}{|c}{$(0,3,0)$} & \multicolumn{1}{|l}{\textrm{U}%
$(0,\theta)$} & \multicolumn{1}{|c}{$\theta^{-1}$} &
\multicolumn{1}{|c|}{$(0,3)$}\\\hline
\multicolumn{1}{|l}{\textrm{HN}$(0,\sigma)$} & \multicolumn{1}{|c}{$\mathbf{1}%
(0<\sigma<10^{2})$} & \multicolumn{1}{|c}{$(0,3)$} &
\multicolumn{1}{|l}{\textrm{We}$(\alpha,\beta)$} &
\multicolumn{1}{|c}{$\mathbf{1}(0\leq\alpha,\beta\leq10^{2})$} &
\multicolumn{1}{|c|}{$(3,1)$}\\\hline
\multicolumn{1}{|l}{\textrm{IGau}$(\mu,\lambda)$} &
\multicolumn{1}{|c}{$\mathbf{1}(0\leq\mu,\lambda\leq10^{2})$} &
\multicolumn{1}{|c}{$(3,1)$} & \multicolumn{1}{|l}{} & \multicolumn{1}{|l}{} &
\multicolumn{1}{|c|}{}\\\hline\hline
\multicolumn{6}{l}{\textit{Abbreviations for distributions:} \ \textrm{BS}
Birnbaum Saunders; \textrm{Exp} exponential; \textrm{Ga} gamma;}\\
\multicolumn{6}{l}{\textrm{Geom} geometric; \textrm{GEV} generalized extreme
value; \textrm{HN} half normal; \textrm{IGau}\ inverse Gaussian;}\\
\multicolumn{6}{l}{\textrm{LN}\ log normal; \textrm{NB }negative binomial;
\textrm{Poi} Poisson; \textrm{U} uniform; \textrm{We} Weibull.}\\
&  & \multicolumn{1}{l}{} &  &  & \multicolumn{1}{l}{}%
\end{tabular}%
\normalsize

\end{center}

For each of these 19 models, for a given data set $\mathcal{Y}_{n}$, the exact
likelihood is tractable and has the form $f_{\boldsymbol{\theta}}%
(\mathcal{Y}_{n})=%
{\textstyle\prod\nolimits_{i=1}^{n}}
f_{\boldsymbol{\theta}}(y_{i})$, with corresponding cdf $F_{\boldsymbol{\theta
}}$ and quantile function $q_{\boldsymbol{\theta}}(\lambda
)=F_{\boldsymbol{\theta}}^{-1}(\lambda)$. For example, the Student's
$\mathrm{T}(y\mid\mu,\sigma,\nu)$ distribution is defined by cdf (\ref{tcdf})
and corresponding likelihood pdf (\ref{tpdf}) with parameters
$\boldsymbol{\theta}=(\mu,\sigma,\nu)$. Its quantile function
$q_{\boldsymbol{\theta}}(\lambda)=F_{\boldsymbol{\theta}}^{-1}(\lambda
)=\mathrm{T}^{-1}(\lambda\mid\mu,\sigma,\nu)$ either has an explicit form or
can be solved by a simple power series approximation, depending on the value
of the degrees of freedom parameter $\nu>0$
\citep[][]{Shaw06}%
. This model was assigned a rather uninformative prior distribution pdf
$\pi(\boldsymbol{\theta})$ shown in Table 1. For this model, the QIL\ is
constructed from a subset of $d(\epsilon)\leq n$ chosen quantiles
$\widehat{\mathbf{q}}_{n,d}=(\widehat{q}_{j}=\widehat{q}(\lambda_{j}%
))_{j=1}^{d}$ of the given data set $\mathcal{Y}_{n}$, and from the vector
quantiles $\mathbf{q}_{\boldsymbol{\theta},d}=(q_{\boldsymbol{\theta}}%
(\lambda_{j})=\mathrm{T}^{-1}(\lambda_{j}\mid\mu,\sigma,\nu))_{j=1}^{d}$, on
$d$ equally-spaced Student cdf probabilities $\boldsymbol{\lambda}_{d}%
=(\tfrac{j}{n+1})_{j=1}^{d}$ in $(0,1)$. All of these steps for the Student
model are performed similarly for each of the other 18 models.

The design of the simulation study is as follows. For each of the 19 models, a
data set $\mathcal{Y}_{n}=\{y_{i}\}_{i=1}^{n}$ was simulated as iid from the
model likelihood distribution $f_{\boldsymbol{\theta}}(y)$ with given true
data generating parameters (mentioned in Table 1), for each of three sample
sizes $n=200,$ $2\,000$, and $20\,000$, and for each of three number of
quantiles $d(\epsilon)$ for QIL, given by $d(.001)$, $d(.01)$, or $d(.1)$.
Then, for each simulated data set, and corresponding model, and corresponding
condition defined by $(n$, $d(\epsilon))$, maximum likelihood estimates
(MLEs)\ of the model parameters were computed from the exact model likelihood
using all the $n$ data points. From the same data set, the QIL-based MAP and
posterior covariance matrix of the model parameters (under a given number of
quantiles $d(\epsilon)$) were estimated by the penalized least-squares
estimation (PLS)\ procedure that used the MLEs as starting parameter values.
For each of the discrete \textrm{Bernoulli}$(\theta)$, \textrm{Geom}$(\theta
)$, \textrm{NB}$(\theta,r)$, and \textrm{Poi}$(\theta)$ models, this
estimation was more conveniently undertaken by computing QIL from the normal
quantiles $q_{\theta}(\lambda)=\mathrm{N}^{-1}(\lambda\,|\,\mu_{\theta}%
,\sigma_{\theta}^{2})$ and the sample normal quantiles $\widehat{q}%
(\lambda)=\mathrm{N}^{-1}(\lambda\,|\,\widehat{\mu},\widehat{\sigma}^{2})$.
Here, $(\mu_{\boldsymbol{\theta}},\sigma_{\boldsymbol{\theta}}^{2})$ are the
mean and variance of $Y$ conditionally on parameters $\boldsymbol{\theta}$ of
the given discrete model, and $(\widehat{\mu},\widehat{\sigma}^{2})$ are the
sample mean and variance of the given simulated data set, $\mathcal{Y}_{n}$.

From the simulated data, we computed the Root Mean Squared Error (RMSE)\ for
the MLEs, and for the QIL MAP\ estimates. In each case, the RMSE is the root
of the average squared difference between the true and given estimate,
averaged over all model parameters, over all models, and over all simulated
data sets, for each of the nine simulation conditions defined by a specific
value of $(n$, $d(\epsilon))$.

The RMSEs were computed by three different choices of $d(\epsilon)$, i.e., for
$\epsilon=.001$, $\epsilon=.01$, and $\epsilon=.1$, in order to evaluate the
sensitivity of QIL-based Bayesian posterior inferences over different choices
of the number of quantiles $d(\epsilon)$. Also, for each of the nine
conditions defined by give values $(n$, $d(\epsilon))$, we computed the total
computation time.

The top row of Figure 1 presents the RMSE results of this simulation study.
The results show that over all sample size conditions $n=200$, $2\,000$, and
$20\,000$, and over all the different choices of $d(.001)$, $d(.01)$, and
$d(.1)$ for QIL, the RMSEs of QIL\ MAP were quite similar to the respective
RMSEs for MLE, especially for larger sample sizes. This is a confirming result
for QIL.\ This is because according to standard asymptotic theory, the
MAP\ estimate converges to the MLE\ estimate as the sample size grows large,
especially given the rather uninformative priors assigned for the 19 Bayesian
models (resp.).

The RMSEs of the QIL MAP\ estimates were very similar over the different
choices of $d(.001)$, $d(.01)$, and $d(.1)$, suggesting that QIL\ can be
insensitive to the chosen number of quantiles $d(\epsilon)$ for small
$\epsilon$. Over all models and simulated data sets, and within each of the
nine simulation conditions, the median $d(.001)$ was $200$ for $n=200$; $1333$
for $n=2\,000$; $1025$ for $n=20\,000;$ the median $d(.01)$ was $133$ for
$n=200$; $102$ for $n=2\,000$, $100$ for $n=20\,000$; and the median for
$d(.1)$ was $9$ for $n=200,$ $2\,000$, and for $20\,000$.

The bottom row of Figure 1 presents the computation times, which took only a
few sections for each simulation condition $(n$, $d(\epsilon))$. Naturally,
the time was shown to be a monotone increasing function of $d(\epsilon)$, the
number of quantiles for QIL. The results of the simulation study suggests that
$d(.01)$ provides the best default choice, considering both the RMSE\ and
computation time of QIL. Therefore, in the sequel, QIL\ will often be
constructed from $d(.01)$ or all $d(0)=n$ quantiles.

\begin{center}
--- Insert Figure 1 here (figure at end of paper) ---
\end{center}

Next, to further illustrate QIL, we analyzed the \texttt{Sulfur} data set
using the Bayesian normal \textrm{N}$(\mu,\sigma^{2})$ model, assigned a
conjugate normal inverse-gamma prior $\pi(\boldsymbol{\theta})\propto\exp
[-\mu^{2}/(2\sigma^{2}10^{2})-\sigma^{-2}]$. This data set, available in the R
software package \texttt{Openair}
\citep{CarslawRopkins12}%
, contains $n=65\,533$ observations of the pollutant sulfur dioxide, measured
in ppb/100 concentration. They were obtained from hourly measurements
collected at the Marylebone (London) air quality monitoring supersite between
1st January 1998 and 23rd June 2005.

For the normal model based on the exact normal likelihood, the marginal
posterior distributions for $(\mu,\sigma^{2})$ can be directly computed as a
normal distribution for $\mu$, and a gamma distribution for $\sigma^{-2}$
\citep[][p.440]{BernardoSmith94}%
. For comparison purposes, we estimated the posterior distribution of the
normal model based on QIL\ using $d(.01)=96$ quantiles, using $50\,000$
sampling iterations of the Adaptive Metropolis (AM)\ algorithm ($\S 2.3.2$%
).\ The algorithm, which used the sample mean and variance as starting values,
completed in $32$ seconds. As Appendix Figure 1A (top) shows, the marginal
posterior distributions of $(\mu,\sigma^{2})$ based on the exact likelihood
nearly match those based on QIL.

\subsection{$g$-and-$h$ and $g$-and-$k$ Distributions for Univariate iid Data}

The $g$-and-$h$
\citep{Tukey77}
and $g$-and-$k$ distributions
\citep{MacGillivray92}
each extends the normal distribution by allowing for more skewness or heavier
(or lighter) tails, and can fit a wide variety distribution shapes with four
interpretable parameters. Such a distribution is defined by a quantile
function via some transformation $Y=A+BG(Z)H(Z)$ of the standard normal
variable $Z\sim\mathrm{N}(0,1)$, where $A$ is a location parameter, $B$ is a
scale parameter, $G(\cdot)$ introduces asymmetry, and $H(\cdot)$ elongates the
tails of the distribution.

The generalized $g$-and-$h$ and $g$-and-$k$ distributions
\citep{MacGillivray92}%
, respectively, are defined by the quantile functions:
\begin{subequations}
\label{ghgk}%
\begin{align}
q_{\boldsymbol{\theta}}^{(gk)}(u)  &  =F_{gh}^{-1}(u;A,B,g,h)=A+B(1+c\tanh
[(g/2)z_{u}])z_{u}\exp[(h/2)z_{u}^{2}],\label{gh}\\
q_{\boldsymbol{\theta}}^{(gh)}(u)  &  =F_{gk}^{-1}(u;A,B,g,k)=A+B(1+c\tanh
[(g/2)z_{u}])z_{u}(1+z_{u}^{2})^{k}, \label{gk}%
\end{align}
where $z_{u}=\mathrm{N}^{-1}(u\boldsymbol{\mid}0,1)$ is the standard normal
quantile function. The $g$-and-$h$ model has parameters $\boldsymbol{\theta
}=(A,B,g,h)$, while the $g$-and-$k$ model has parameters $\boldsymbol{\theta
}=(A,B,g,k)$.\ Here, $g$ controls skewness, $h$ or $k$ is the kurtosis (tail
size) added to the $\mathrm{N}(0,1)$ distribution, and $c=.8$ is the standard
choice of overall asymmetry constant
\citep{RaynerMacGillivray02,MacGillivray86}%
. A proper distribution is guaranteed by the parameter values $A\in%
\mathbb{R}
$, $B>0$, $h\geq0$ or $k\geq0$, and $0\leq c<c^{\ast}\approx.83$
\citep{RaynerMacGillivray02}%
.

The $g$-and-$h$ and $g$-and-$k$ distributions (resp.) each do not generally
admit a closed-form expression for its exact likelihood pdf. Instead, its
exact likelihood is expressible in terms of derivatives of quantile functions,
and needs to be computed completely numerically for each of the individual
data points $y_{i}$
\citep{RaynerMacGillivray02,Prangle17}%
. This likelihood computation method is slow when either the data sample size
$n$ is large, and hundreds of time slower than computing the normal pdf
\citep{RaynerMacGillivray02,Prangle17}%
. Further, for either the $g$-and-$h$ or the $g$-and-$k$ model, the exact
likelihood pdf can be highly multimodal when the sample size does not greatly
exceed 100, or when the data exhibits extreme non-normality
\citep{RaynerMacGillivray02}%
. Such computational challenges have urged considerable methodological work on
likelihood-based inference for both models
\citep[][and references therein]{Prangle17}%
.

The QIL\ approach to approximate Bayesian posterior inference is now
illustrated for the $g$-and-$h$ and $g$-and-$k$ models. For each model, a
rather uninformative prior distribution was assigned to the parameters, given
by $\pi(\boldsymbol{\theta})\propto\mathbf{1}(-10\leq A,g\leq10,0\leq B,h$ or
$k\leq10)$. For each model, an iid data set $\mathcal{Y}_{n}$ of size
$n=20\,000$ was simulated under the model based the parameters $A=-.7,$
$B=1.7$, $g=-.4,$ $h$ or $k=.5$; and the QIL\ was constructed from
$d(.01)=100$ quantiles $\widehat{\mathbf{q}}_{n,d}=(\widehat{q}_{j}%
=\widehat{q}(\lambda_{j}))_{j=1}^{d}$ of the data set $\mathcal{Y}_{n}$, and
from the vector quantiles $\mathbf{q}_{\boldsymbol{\theta},d}%
=(q_{\boldsymbol{\theta}}(\lambda_{j}))_{j=1}^{d}$ on cdf probabilities
$\boldsymbol{\lambda}_{d}=(\tfrac{j}{n+1})_{j=1}^{d}$ in $(0,1)$. Here, the
quantile function $q_{\boldsymbol{\theta}}(\lambda)$ for the $g$-and-$h$ model
is given by (\ref{gh}), and the quantile function for the $g$-and-$k$ model is
given by (\ref{gk}).

For each model and respective data set, the QIL\ MAP estimate
$\widehat{\boldsymbol{\theta}}_{Q}$ was calculated as the solution that
attained the minimum PLS according to (\ref{QLEpen2}), among all solutions
(resp.)\ obtained from separate runs of the PLS algorithm that used $232$
different starting values of $(A,B,g,h$ or $k)$ (resp.). One starting value is
given by a simple plug-in approximate estimator
\citep{Hoaglin85}
of the parameters $\boldsymbol{\theta}=(A,B,g,h)$ for the ungeneralized
$g$-and-$h$ distribution
\citep{Tukey77}%
. The other $231$ starting values were, in combination, defined by starting
values of $A$ as $-1$, $0$, $1$, or the data median; of $B$ as $1$, $10$,
$100$, or half the data interquartile range; of $g$ as $-10$,$-1$, $0$, $1$,
or $10$; and of $h$ or $k$ as $0$, $1$, or $10$
\citep[e.g.,][p.61]{RaynerMacGillivray02}%
.

For each of the $g$-and-$h$ model and the $g$-and-$k$ model, and their
simulated data sets (resp.), the RMSE\ of the QIL\ MAP estimate was calculated
as the root of the average squared difference between the true and
MAP\ estimate, averaged over all model parameters. It was found that the
RMSE\ was $.03$ for the $g$-and-$h$ model, and was $5.68$ for the $g$-and-$k$
model. We compared with the RMSE of point estimates using the exact model
likelihood, and using the finite differential stochastic approximation
(fdsa)\ method to estimating the MLEs
\citep{Prangle17}%
. For the $g$-and-$h$ model, this method reached no solution after 1000
iterations; while this method obtained a RMSE\ of $2.72$ for the $g$-and-$k$ model.

Next, for each of the $g$-and-$h$ and $g$-and-$k$ models, we estimated the
posterior distribution of the model parameters, respectively, by running
$10^{5}/2$ and $10^{6}$ sampling iterations of the adaptive Metropolis
algorithm ($\S 2.3.2$), using the QIL\ MAP estimates as algorithm starting
values. For each model, the RMSE\ was computed as the root of the average
squared difference between the true parameters and corresponding sampled
parameters, averaging over all model parameters and sampling iterations. The
RMSE\ was $.16$ for the $g$-and-$h$ model, and was $3.93$ for the $g$-and-$k$
model. In comparison, for the ABC\ method (ABCo) based on octiles of the data
\citep{Prangle17}%
, the RMSE\ was $1.79$ for the $g$-and-$h$ model, and was $2.48$ for the
$g$-and-$k$ model, based on $10^{6}$ Monte Carlo samples from the given model
prior (and then retaining for analysis $1000$ samples that produced sampled
data sets with the smallest distances to the given data set). For the
ABC\ method (ABCa) based on using all of the data points
\citep{Prangle17}%
, the RMSE\ was $3.61$ for the $g$-and-$h$ model, and $3.61$ for the
$g$-and-$k$ model, based on $10^{5}/2$ Monte Carlo samples from the given
model prior (and then retaining $1000$ samples with the smallest distances).
Finally, using the exact model likelihood, attempts were made to estimate the
full posterior distribution of each model, using the adaptive Metropolis (AM)
algorithm
\citep{Prangle17}%
, and using a brut-force approach to the synthetic likelihood method
\citep{TurnerSederberg14}%
. Both of these methods were too computationally expensive to implement on the
simulated data sets. In conclusion, the QIL\ method attained better or
competitive RMSE, compared to ABCo and ABCa.
\end{subequations}
\begin{center}%
\footnotesize
\begin{tabular}
[c]{ccccccc}%
\multicolumn{7}{l}{\textbf{Table 2.} Computation times from real data and
simulated data,}\\
\multicolumn{7}{l}{for the $g$-and-$h$ and $g$-and-$k$ models.}\\\hline\hline
Model & Likelihood & Algorithm & iterations & Data Set & $n$ & time\\\hline
$g$-and-$h$ & QIL & PLS & NA & \texttt{Simulated} & $20\,000$ & $21$ secs\\
& exact & fdsa & $1\,000$ & \texttt{Simulated} & \multicolumn{2}{c}{No
solution reached}\\
& QIL & AM & $10^{5}/2$ & \texttt{Simulated} &  & $61.2$ secs\\
& ABCo & Rejection & $10^{6}$ & \texttt{Simulated} &  & $58$ secs\\
& ABCa & Rejection & $10^{5}/2$ & \texttt{Simulated} &  & $5.2$
mins\\\cline{2-7}
& QIL & PLS & NA & \texttt{Sulfur} & $65\,533$ & $19$ secs\\
& exact & fdsa & $1\,000$ & \texttt{Sulfur} & $65\,533$ & $3.1$ mins\\
& QIL & AM & $10^{6}/2$ & \texttt{Sulfur} & $65\,533$ & $6.8$ min\\
& ABCo & Rejection & $10^{6}/2$ & \texttt{Sulfur} & $65\,533$ & $29$
secs\\\hline
$g$-and-$k$ & QIL & PLS & NA & \texttt{Simulated} & $20\,000$ & $24$ secs\\
& exact & fdsa & $1\,000$ & \texttt{Simulated} &  & $3.4$ mins\\
& QIL & AM & $10^{6}$ & \texttt{Simulated} &  & $19.2$ mins\\
& ABCo & Rejection & $10^{6}$ & \texttt{Simulated} &  & $58$ secs\\
& ABCa & Rejection & $10^{5}/2$ & \texttt{Simulated} &  & $5.8$
mins\\\cline{2-7}
& QIL & PLS & NA & \texttt{Sulfur} & $65\,533$ & $24$ secs\\
& exact & fdsa & $1\,000$ & \texttt{Sulfur} & $65\,533$ & $3.9$ mins\\
& QIL & AM & $10^{6}/2$ & \texttt{Sulfur} & $65\,533$ & $6.5$ mins\\
& ABCo & Rejection & $10^{6}/2$ & \texttt{Sulfur} & $65\,533$ & $28$
secs\\\hline\hline
\multicolumn{7}{l}{}%
\end{tabular}%
\normalsize

\end{center}

To further illustrate, the \texttt{Sulfur Dioxide} data set of $n=65\,533$
observations was analyzed by the $g$-and-$h$ model, and by the $g$-and-$k$
models. For each model, the posterior distribution of the model parameters was
estimated based on QIL\ using $d(.01)=100$ quantiles and $10^{6}/2$ sampling
iterations of the adaptive Metropolis algorithm, and by the ABCo method using
$10^{6}/2$ sampling iterations (and then basing the posterior results on the
$1000$ samples that attained the smallest distances). For the $g$-and-$h$
model, the prior $\pi(\boldsymbol{\theta})\propto\mathbf{1}(-10\leq
A,g\leq10,0\leq B,h$ or $k\leq10)$ was specified for the QIL, and for the ABCo
methods. For the $g$-and-$k$ model, the same prior was employed for ABCo. But
for this same model, the QIL method was based on the flat uninformative
$\pi(\boldsymbol{\theta})\propto1$ prior, which is possible to use under the
adaptive Metropolis (AM) algorithm. The ABCa, exact likelihood with adaptive
Metropolis, and Brut methods were each too computationally prohibitive to
analyze the large \texttt{Sulfur Dioxide} data set. Appendix Figure A1
presents, for the \texttt{Sulfur} data, the marginal posterior distribution
estimates of the $g$-and-$h$ and $g$-and-$k$ models under QIL (resp.).

For the simulated and \texttt{Sulfur Dioxide} data sets, Table 2 shows that
the QIL\ method was better or at least competitive versus ABCo and ABCa in
computation time. QIL do not require the selection of many tuning parameters,
unlike the ABC methods. For ABC practice, such a selection may involve trying
different values of the tuning parameters with the aim of optimizing them, but
this adds to ABC's computation time.

\subsection{Logistic Regression for Univariate non-iid Data}

A standard Bayesian binary regression model is defined by:
\begin{subequations}
\label{BinaryReg}%
\begin{align}
f_{\boldsymbol{\theta}}(\mathcal{Y}_{n})  &  =%
{\displaystyle\prod\limits_{i=1}^{n}}
[G(\mathbf{x}_{i}^{\mathbf{\intercal}}\boldsymbol{\beta})]^{y_{i}%
})[1-G(\mathbf{x}_{i}^{\mathbf{\intercal}}\boldsymbol{\beta})]^{1-y_{i}}\\
\pi(\boldsymbol{\beta},\boldsymbol{\eta})  &  =\pi(\beta_{1},\ldots,\beta
_{p}\boldsymbol{\mid\eta})\pi(\boldsymbol{\eta}).
\end{align}
for a given set of data $\mathcal{Y}_{n}=(\mathbf{y}_{n},\mathbf{X}%
_{n})=(y_{i},\mathbf{x}_{i}^{\mathbf{\intercal}})_{n\times(p+2)}$ of binary
dependent responses $\mathbf{y}_{n}=(y_{i})_{n\times1}$ and $p$ covariates
$\mathbf{X}_{n}=(\mathbf{x}_{i}^{\mathbf{\intercal}})_{n\times(p+1)}$, with
$y\in\{0,1\}$, $\mathbf{x}^{\mathbf{\intercal}}=(1,x_{1},\ldots,x_{p})$, and
cdf $G:%
\mathbb{R}
\rightarrow(0,1)$ being the inverse of the chosen link function. The most
popular choice of inverse link $G$ includes the standard \textrm{Logistic}%
$(0,1)$ cdf, $G(\cdot)=\frac{\exp(\cdot)}{1+\exp(\cdot)}$, which defines a
logit model. Another popular inverse link is given by the standard Normal cdf,
$G(\cdot)=$ \textrm{N}$(\cdot\mid0,1)$, which defines a probit model. The
model parameters are given by $\boldsymbol{\theta}=(\boldsymbol{\beta
},\boldsymbol{\eta})$, with coefficient parameters $\boldsymbol{\beta}%
=(\beta_{0},\beta_{1},\ldots,\beta_{p})^{\intercal}$, hyperparameter
$\boldsymbol{\eta}$, and a flat prior $\pi(\beta_{0})\propto1$ for the
intercept parameter. Setting $\boldsymbol{\eta}$ to a fixed value corresponds
to a degenerate prior distribution $\pi(\boldsymbol{\eta})$ which assigns mass
1 at $\boldsymbol{\eta}$.

For inference of the model's exact posterior distribution $\pi
(\boldsymbol{\beta},\boldsymbol{\eta}\mid\mathcal{Y}_{n})$, standard MCMC
algorithms include the P\'{o}lya-gamma (P-G) sampler for the logit model
\citep{PolsonEtAl13}%
, and the Gibbs sampler for the probit model
\citep{AlbertChib93}%
. Such an algorithm, in each sampling iteration, draws $n$ latent variables of
the observed binary responses $(y_{i})_{n\times1}$ (resp.) from their
respective conditional distributions given $\boldsymbol{\beta}$, and draws
$\boldsymbol{\beta}$ from its full conditional distribution conditionally on
the updated latent variables. Updating the latent variables is computationally
costly when the sample size $n$ is large. The same iteration also requires
inverting a $(p+1)\times(p+1)$ precision matrix if $\boldsymbol{\eta}$ is
treated as an unknown parameter and assigned a non-degenerate prior
$\pi(\boldsymbol{\eta})$, a step that is costly when the number of covariates
$p$ is large.

We now explain how to specify the QIL as an approximate likelihood for the
Bayesian binary regression model, for the purposes of increasing computational
speed in the estimation of the posterior distribution. Here, the QIL is given
by equation (\ref{QILnoniid}), with pivotal statistics $t_{\boldsymbol{\theta
}}(y_{i})$ given by (\ref{noniidPivotal}) for $i=1,\ldots,n$, where the sample
size equals the number of groups with $K=n$ (so that $i=k$ for $k=1,\ldots,K$,
and $d_{i}=1$ and $\lambda_{i}=1/2$\ for $i=1,\ldots,n=K$). These $n$ pivotal
statistics have independent asymptotic chi-square distributions,
$\{t_{\boldsymbol{\theta}}(y_{i})\}_{i=1}^{n}\overset{\text{ind}}{\sim}%
\chi_{1}^{2}$. Further, the QIL\ pivotal statistics $t_{\boldsymbol{\theta}%
}(y_{i})$ provides a support vector machine approach to classification using
hinge loss, and can be applied to any choice of smooth inverse link function
$G$. Using algebra, each pivotal statistic $t_{\boldsymbol{\theta}}(y_{i})$
can be derived for the binary regression model as:%
\end{subequations}
\begin{equation}
t_{\boldsymbol{\theta}}(\mathcal{Y}_{n_{k}})=t_{\boldsymbol{\theta}}%
(y_{i})=\dfrac{\mathbf{1}(y_{i}\neq\mathbf{1}(G(\mathbf{x}_{i}%
^{\mathbf{\intercal}}\boldsymbol{\beta})\geq1/2))}{.25/[\max\{G(\mathbf{x}%
_{i}^{\mathbf{\intercal}}\boldsymbol{\beta}),1-G(\mathbf{x}_{i}%
^{\mathbf{\intercal}}\boldsymbol{\beta})\}]^{2}}, \label{Qhinge}%
\end{equation}
which equals the squared hinge loss $[1-(-1)^{1-y_{i}}(2G(\mathbf{x}%
_{i}^{\mathbf{\intercal}}\boldsymbol{\beta})-1)]^{2}$ for an incorrect
classification ($y_{i}\neq\mathbf{1}(G(\mathbf{x}_{i}^{\mathbf{\intercal}%
}\boldsymbol{\beta})\geq1/2$), and equals zero for a correct classification
\citep[][Table 12.1]{HastieTibsFriedman09}%
.

Then, for the Bayesian logistic model with multivariate normal prior pdf
$\pi(\boldsymbol{\beta\mid\eta})=\mathrm{n}(\beta_{1},\ldots,\beta_{p}$
$\boldsymbol{\mid0},\boldsymbol{T}_{\boldsymbol{\eta}})$ and fixed
hyperparameter $\boldsymbol{\eta}$, and for the given set of data
$\mathcal{Y}_{n}=(y_{i},\mathbf{x}_{i}^{\mathbf{\intercal}})_{n\times(p+2)}$,
the QIL\ MAP and posterior covariance matrix estimates of the coefficient
parameters can be found by the penalized maximum likelihood estimation
algorithm. Since the QIL is defined by degrees of freedom parameters $d_{i}%
=1$, the posterior mode estimate of $\boldsymbol{\beta}$ is also a minimum
chi-square estimate. Also, the Adaptive Metropolis (AM) sampling algorithm
($\S 2.3.2$) can be used to estimate quantities from the posterior
distribution for this model. For either algorithm, the MLE of
$\boldsymbol{\beta}$ can be used as the starting values for the regression
coefficient parameters.

We now consider a simulation study of the QIL method, mainly using all
$d(0)=n$ quantiles of the full data $\{(y_{i},\mathbf{x}_{i}^{\intercal
})\}_{i=1}^{n}$, unless otherwise indicated. Each simulated data set was based
on either sample size $n=30\,000,$ $10^{5}$, $10^{6}$, or $10^{7}$, and $p=8$
covariates with data generating parameters $\boldsymbol{\beta}%
=(0,3,1.5,0,0,2,0,0,0)^{\intercal}$; or sample size $n=30\,000,$ $10^{5}$, or
$10^{6}$, and $p=100$ covariates with data generating parameters
$\boldsymbol{\beta}=(0,3,1.5,0,0,2,\ldots,3,1.5,0,0,2)^{\intercal}$. Each data
set $\mathcal{Y}_{n}=(y_{i},\mathbf{x}_{i}^{\mathbf{\intercal}})_{n\times
(p+2)}$ was simulated by first drawing multivariate normal samples
$\{\mathbf{x}_{(1:p),i}\}_{i=1}^{n}\overset{\text{iid}}{\sim}$ \textrm{N}%
$_{p}(\mathbf{0},\Sigma)$, with $\Sigma=(.5^{|j-l|})_{p\times p}$ for distinct
covariates $j,k\in\{1,\ldots,p\}$, and then drawing $y_{i}\mid\mathbf{x}%
_{i}\overset{\text{ind}}{\sim}$ \textrm{Bernoulli}$(G(\mathbf{x}%
_{i}^{\mathbf{\intercal}}\boldsymbol{\beta}))$ with $\mathbf{x}_{i}%
^{\intercal}=(1,\mathbf{x}_{(1:p),i}^{\intercal})$ for $i=1,\ldots,n$. For the
simulated data set $\{(y_{i},\mathbf{x}_{i}^{\intercal})\}_{i=1}^{n}$ with
sample size $n=10^{7}$ and $p=100$ covariates, we also considered analyzing
the data subset formed by $d(.001)=1\,001$ sample quantiles of the Mahalanobis
depths of the $(y_{i},\mathbf{x}_{i}^{\intercal})$ samples, respectively (see
$\S 2.2$).

\begin{center}
--- Insert Figure 2 here (figure at end of paper) ---
\end{center}

First, for each simulated data set $\mathcal{Y}_{n}$, we obtained the QIL
MAP\ and posterior covariance estimates of the coefficients $\boldsymbol{\beta
}$ for the logistic model based on a rather uninformative $p$-variate normal
prior pdf, given by $\pi(\boldsymbol{\beta\mid\eta})=\mathrm{n}_{p}%
(\boldsymbol{\beta}_{1:p}\mid\mathbf{0},10^{8}\mathbf{I}_{p})$. The left
panels of Figure 2\ present the RMSE of the QIL MAP estimate for each
simulated data set. Here, each RMSE is the root of the average squared
difference between the QIL\ MAP\ estimate and the true data-generating
coefficient parameter, averaging over all coefficient parameters. Figure 2
also presents the minimum, median and maximum absolute differences between the
QIL MAP and MLE\ estimates over all coefficient parameters (see Figure 2,
middle panels); and between the QIL\ posterior standard deviations (qilSD) and
standard errors (mleSE) (see Figure 2, right panels). Again, this is a
validating result for QIL. Generally speaking the MAP\ and MLE\ estimates
converges towards equality as the sample size grows large, especially given
the rather uninformative prior assigned to the coefficients.

For further illustration, we consider a binary logit regression analysis of
the \texttt{Diabetes} data set, containing the medical record information of
$n=101\,766$ hospital patients, obtained from:\newline%
\url
{https://archive.ics.uci.edu/ml/datasets/Diabetes 130-US hospitals for years 1999-2008}%
.\newline These data were used to study the relationship between an early
hospital readmission indicator, and $p=27$ covariates defined by binary ($0$
or $1$)\ indicators of discharge type, race, gender, admission source, medical
specialty, age category, primary diagnosis, and glucose control levels (HbA1c)
interacted with an indicator of change in diabetes medications
\citep[][Table 4]{StrackEtAl14}%
.

\begin{center}%
\footnotesize
\begin{tabular}
[c]{cccccccc}%
\multicolumn{8}{l}{\textbf{Table 3.} Computation times from real data and
simulation data, logit and probit models.}\\\hline\hline
Model & Likelihood & Algorithm & iterations & Data Set & $n$ & $p$ &
time\\\hline
Logit & QIL & PLM & NA & Simulated & $30\,000$ & $8$ & $4$ secs\\
Logit & QIL & AM & $10^{5}$ & Simulated & $30\,000$ & $8$ & $4$ mins\\
Logit & Exact & P-G & $10^{5}$ & Simulated & $30\,000$ & $8$ & $40.1$ mins\\
Probit & Exact & PrG & $10^{5}$ & Simulated & $30\,000$ & $8$ & $5.8$
mins\\\hline
Logit & QIL & PLM & NA & Simulated & $100\,000$ & $8$ & $18$ secs\\
Logit & QIL & AM & $10^{5}$ & Simulated & $100\,000$ & $8$ & $11.6$ mins\\
Logit & Exact & P-G & $10^{5}$ & Simulated & $100\,000$ & $8$ & $2.2$ hrs\\
Probit & Exact & PrG & $10^{5}$ & Simulated & $100\,000$ & $8$ & $45.8$
mins\\\hline
Logit & QIL & PLM & NA & Simulated & $1\,000\,000$ & $8$ & $3.5$ mins\\
Logit & QIL & AM & $10^{5}$ & Simulated & $1\,000\,000$ & $8$ & $1.9$ hrs\\
Logit & Exact & P-G & $10^{5}$ & Simulated & $1\,000\,000$ & $8$ & $>>3$ hrs\\
Probit & Exact & PrG & $10^{5}$ & Simulated & $1\,000\,000$ & $8$ & $4.7$
hrs\\\hline
Logit & QIL & PLM & NA & Simulated & $10\,000\,000$ & $8$ & $21$ mins\\
Logit & QIL & AM & $10^{5}$ & Simulated & $10\,000\,000$ & $8$ & $19.4$ hrs\\
Logit & Exact & P-G & $10^{5}$ & Simulated & $10\,000\,000$ & $8$ & $>>3$
hrs\\
Probit & Exact & PrG & $10^{5}$ & Simulated & $10\,000\,000$ & $8$ & $>>5$
hrs\\\hline
Logit & QIL & PLM & NA & Simulated & $1\,001$ selected & $8$ & $<1$ sec\\
Logit & QIL & AM & $10^{5}$ & Simulated & $1\,001$ selected & $8$ & $38$
secs\\
Logit & Exact & P-G & $10^{5}$ & Simulated & $1\,001$ selected & $8$ & $1.2$
mins\\
Probit & Exact & PrG & $10^{5}$ & Simulated & $1\,001$ selected & $8$ & $9.3$
secs\\\hline
Logit & QIL & PLM & NA & Simulated & $30\,000$ & $100$ & $1.7$ mins\\
Logit & QIL & AM & $10^{5}$ & Simulated & $30\,000$ & $100$ & $\ 8.4$ mins\\
Logit & Exact & P-G & $10^{5}$ & Simulated & $30\,000$ & $100$ & $13.7$ hrs\\
Probit & Exact & PrG & $10^{5}$ & Simulated & $30\,000$ & $100$ & $2.6$
hrs\\\hline
Logit & QIL & PLM & NA & Simulated & $100\,000$ & $100$ & $5.3$ mins\\
Logit & QIL & AM & $10^{5}$ & Simulated & $100\,000$ & $100$ & $\ 24.8$ mins\\
Logit & Exact & P-G & $10^{5}$ & Simulated & $100\,000$ & $100$ & $>>14$ hrs\\
Probit & Exact & PrG & $10^{5}$ & Simulated & $100\,000$ & $100$ & $>>3$
hrs\\\hline
Logit & QIL & PLM & NA & Simulated & $1\,000\,000$ & $100$ & $1.1$ hrs\\
Logit & QIL & AM & $10^{5}$ & Simulated & $1\,000\,000$ & $100$ & $4.7\ $hrs\\
Logit & Exact & P-G & $10^{5}$ & Simulated & $1\,000\,000$ & $100$ & $>>14$
hrs\\
Probit & Exact & PrG & $10^{5}$ & Simulated & $1\,000\,000$ & $100$ & $>>3$
hrs\\\hline
Logit & QIL & PLM & NA & \texttt{Diabetes} & $101\,766$ & $27$ & $65.4$ secs\\
Logit & QIL & AM & $10^{5}$ & \texttt{Diabetes} & $101\,766$ & $27$ & $13.8$
mins\\
Logit & Exact & P-G & $10^{5}$ & \texttt{Diabetes} & $101\,766$ & $27$ & $1.8$
hrs\\
Probit & Exact & PrG & $10^{5}$ & \texttt{Diabetes} & $101\,766$ & $27$ &
$1.2$ hrs\\\hline
Logit & QIL & AM & $10^{5}$ & \texttt{Florence} & $240$ & $2$ & $1.5$ mins\\
Logit & QIL & AM & $10^{5}$ & \texttt{Florence} & $240$ & $5$ & $1.5$
mins\\\hline\hline
\multicolumn{8}{l}{\textit{Notes: }$\ n$ is sample size, $p$ is dimension of
$Y$ (or $\boldsymbol{Y}$), and $p$ is the number of covariates.}\\
\multicolumn{8}{l}{For Florence data, the logit model with $5$ covariates is
based on a quadratic model.}\\
\multicolumn{8}{l}{`$1\,001$ selected' refers to $d(.001)=1\,001$ multivariate
quantiles selected from the}\\
\multicolumn{8}{l}{10 million observations of $\{(y_{i},\mathbf{x}%
_{i})\}_{i=1}^{n}$. In all other cases involving QIL, $d=n$.}%
\end{tabular}%
\normalsize

\end{center}

The QIL\ was constructed for the logit model using all $d(0)=n=101\,766$
observations. For the 27 covariates and constant (1) term, $\mathbf{x}%
=(1,x_{1},\ldots,x_{27})^{\intercal}$, the model's coefficients
$\boldsymbol{\beta}=(\beta_{0},\beta_{1},\ldots,\beta_{27})^{\intercal}$ were
assigned an improper uniform prior for the intercept $\beta_{0}$, and a
LASSO\ prior $\pi(\boldsymbol{\beta},\boldsymbol{\eta})\propto%
{\textstyle\prod\nolimits_{k=1}^{p}}
\exp(-\lambda|\beta_{k}|)\exp(-\lambda)\lambda^{-1/2}$
\citep{Tibshirani96}
including a $\mathrm{gamma}(\lambda\mid1/2,1)$ hyperprior for the parameter
$\lambda$ which controls how much the coefficients are shrunk towards zero.
Then the corresponding marginal posterior distribution $\pi_{\text{Q}%
}(\boldsymbol{\beta}\mid\mathcal{Y}_{n})$ for the slope coefficients
$(\beta_{1},\ldots,\beta_{27})$ will concentrate near zero for any
insignificant covariates. The LASSO\ prior seems appropriate under the belief
that a subset of the covariates provide significant predictors, among the
large number of covariates. Before data analysis, the observations of each of
the $27$ covariates were standardized to have sample mean zero and variance 1,
so that the shrinking parameter $\lambda$ becomes more meaningful. The joint
posterior distribution $\pi_{\text{Q}}(\boldsymbol{\beta},\lambda
\mid\mathcal{Y}_{n})$ of the \texttt{Diabetes} data was estimated by the
adaptive Metropolis algorithm. The algorithm, in each sampling iteration,
employed the adaptive Metropolis step (\S 2.3.2) to update the coefficients
$\boldsymbol{\beta}$; and then employed another adaptive Metropolis step to
update $\lambda$ which aims to yield an optimal acceptance rate of .44 over
iterations using the Metropolis acceptance ratio
\citep{AtchadeRosenthal05}%
.

Appendix Figure A2 (top)\ presents the estimates marginal posterior
distributions of the slope coefficients $(\beta_{1},\ldots,\beta_{27})$, based
on $10^{5}$\ sampling iterations of the adaptive Metropolis algorithm (using
MLE for $\boldsymbol{\beta}$ and 1/2 for $\lambda$ as the starting values).
These coefficients concentrated near zero for most of the covariates,
indicating that they are insignificant predictors. The marginal posterior mean
(SD) of the shrinkage parameter $\lambda$ was $3.66$ ($.00$).

Each simulated data set and the \texttt{Diabetes} data sets was also analyzed
using the logit model estimated under the P\'{o}lya-gamma (P-G) sampler, and
the probit model estimated by the Gibbs sampler, each assigning a $p$-variate
normal prior $\pi(\boldsymbol{\beta},\boldsymbol{\eta})=\mathrm{n}%
_{p}(\boldsymbol{\beta}_{1:p}\mid\mathbf{0},10^{8}\mathbf{I}_{p})$. Table 3
presents shows that in terms of computation time, the QIL-based estimation
methods for the logit model generally outperformed the P\'{o}lya-gamma (P-G)
sampler and the probit Gibbs sampler.

\subsection{Exponential Random\ Graph (ERG) Model for Network\ Data}

Consider a binary network data matrix among $N$ individuals, given by
$\mathcal{Y}_{n}=(y_{ij}:i\neq j,1\leq i,j\leq N)$, where $y_{ij}=1$ indicates
that there is an edge between $i$ and $j$, and $y_{ij}=0$ if no edge. For such
data, a standard model is the Exponential Random graph (ERG) model, which is
defined by the likelihood
\citep[e.g.,][]{CaimoFriel11}%
:%
\begin{equation}
f_{\boldsymbol{\theta}}(\mathcal{Y}_{n})=\left.  \exp\left\{
\boldsymbol{\beta}^{\intercal}\mathbf{x}(\mathcal{Y}_{n})\right\}  \right/
\mathcal{Z}(\boldsymbol{\beta}); \label{ERG}%
\end{equation}
with normalizing constant:%
\begin{equation}
\mathcal{Z}(\boldsymbol{\beta})=%
{\textstyle\sum\limits_{\forall\mathcal{Y}_{n}^{\prime}}^{2^{N(N-1)}}}
\exp\left\{  \boldsymbol{\beta}^{\intercal}\mathbf{x}(\mathcal{Y}_{n}^{\prime
})\right\}  , \label{ERGnorm}%
\end{equation}
and a known covariate vector $\mathbf{x}(\mathcal{Y}_{n})$ of sufficient
statistics describing the network (e.g., the number of edges, degree
statistics, etc.).

The ERG likelihood (\ref{ERG}) is in general intractable because its
normalizing constant (\ref{ERGnorm}) involves a sum of all possible
$2^{N(N-1)}$ binary matrix events, aside from smaller networks consisting of
$N\leq6$ individuals. The ERG\ model does however imply a logit model for the
data $\mathcal{Y}_{n}$, defined by the tractable pseudo-likelihood:%
\begin{equation}
f_{\boldsymbol{\theta}}(\mathcal{Y}_{n})=%
{\displaystyle\prod\limits_{i\neq j}^{N(N-1)}}
\dfrac{\exp[\boldsymbol{\beta}^{\intercal}\{\mathbf{x}(\mathcal{Y}_{n,ij}%
^{+})-\mathbf{x}(\mathcal{Y}_{n,ij}^{-})\}]}{1+\exp[\boldsymbol{\beta
}^{\intercal}\{\mathbf{x}(\mathcal{Y}_{n,ij}^{+})-\mathbf{x}(\mathcal{Y}%
_{n,ij}^{-})\}]}, \label{logitERG}%
\end{equation}
where $\mathcal{Y}_{n,ij}^{+}$ is the matrix $\mathcal{Y}_{n}$ with $y_{ij}%
=1$, and $\mathcal{Y}_{n,ij}^{-}$ is the matrix $\mathcal{Y}_{n}$ with
$y_{ij}=0$
\citep{StraussIkeda90}%
.

To further illustrate, consider the \texttt{Florentine} \texttt{Business}
social network data. This data set contains observations of business network
ties (financial ties such as loans, credits and joint partnerships) among 16
Renaissance Florentine families who were locked in a struggle for political
control of the city of Florence around 1430
\citep[][Table 1]{BreigerPattison86}%
. This data set gives the network ties ($y=1$) among the 16 families, as
follows. Family 3 had ties with families 5, 6, 9, and 11 (resp.); family 4 had
ties with families 7, 8, 11; 5 with 3, 8, 11; 6 with 3, 9; 7 with 4, 8; 8 with
4, 5, 7, 11; 9 with 3, 6, 10, 14, 16; 10 with 9; 11 with 3, 4, 5, 8; 14 with
9; 16 with 9. All other pairs of families have no ties ($y=0$). for the ERG
model we consider two network covariates, $\mathbf{x}(\mathcal{Y}_{n}%
)=(x_{1}(\mathcal{Y}_{n}),x_{2}(\mathcal{Y}_{n}))$, given by the number of
network ties $x_{1}(\mathcal{Y}_{n})=%
{\textstyle\sum\nolimits_{i<j}}
y_{ij}$ and the number of two-stars $x_{2}(\mathcal{Y}_{n})=%
{\textstyle\sum\nolimits_{i<j<k}}
y_{ik}y_{jk}$.

Statistical estimation of the ERG parameters, using either the exact
ERG\ likelihood (\ref{ERG}) or pseudo-likelihood (\ref{logitERG}), was shown
to be problematic in previous research. Monte Carlo Maximum Likelihood
Estimation (MC-MLE) using the exact ERG likelihood (\ref{ERG}) produced, for
the two covariates, MLEs $\widehat{\boldsymbol{\beta}}_{\text{MC-MLE}%
}=(-3.39,.30)$ with unreasonably large standard errors $(21.69,.79)$ (SEs) for
$\boldsymbol{\beta}=(\beta_{1},\beta_{2})^{\intercal}$
\citep{CaimoFriel11}%
. Pseudo-likelihood estimation (MPLE)
\citep{Besag74}%
, which aims to maximize the likelihood (\ref{logitERG}) obtained the
estimates $\widehat{\boldsymbol{\beta}}_{\text{MPLE}}=(-3.39,.35)$, still with
unreasonably large SEs $(.70,.14)$ for the coefficients $\boldsymbol{\beta
}=(\beta_{1},\beta_{2})^{\intercal}$ for the same two covariates. The adaptive
exchange MCMC\ algorithm, which aims to infer the posterior distribution
$\pi(\boldsymbol{\beta\mid}\mathcal{Y}_{n})$ based on the original ERG
likelihood (\ref{ERG}), relied on priors $\beta_{1}\sim\mathrm{Uniform}(-4,0)$
and $\beta_{2}\sim\mathrm{Uniform}(0,8)$ informed by MPLE and SE\ estimates
\citep[][p.565]{JinEtAl13}%
. Therefore, this algorithm produced empirical-Bayes rather than fully-Bayes inferences.

For the same network data set, we considered the logit model as discussed in
the previous subsection, using the QIL specifications for the ERG\ logit
likelihood (\ref{logitERG}). Also, the LASSO\ prior distribution pdf
$\pi(\boldsymbol{\beta},\boldsymbol{\eta})\propto%
{\textstyle\prod\nolimits_{k=1}^{p}}
\exp(-\lambda|\beta_{k}|)\exp(-\lambda)\lambda^{-1/2}$ was assigned to the
coefficients $\boldsymbol{\beta}=(\beta_{1},\beta_{2})^{\intercal}$ of the two
network covariates. Intuitively, the coefficient shrinkage enforced by the
LASSO prior will encourage their marginal posterior standard deviations to be
smaller relative to the SEs for the coefficients reported earlier.

Indeed, this QIL-based posterior analysis of the ERG\ model resulted in, for
the regression coefficients $\boldsymbol{\beta}=(\beta_{1},\beta
_{2})^{\intercal}$ of the two network covariates, marginal posterior means
$\overline{\boldsymbol{\beta}}=(-.07,-.03)^{\intercal}$ and corresponding
marginal posterior standard deviations $(.06,.04)$, the latter which express
smaller and more reasonable posterior uncertainties. These results are based
on $10^{5}$ iteration run of the adaptive Metropolis algorithm (\S 2.3.2).
Appendix Figure A2 (bottom)\ presents the marginal posterior distribution
estimates of the slope coefficients $(\beta_{1},\beta_{2})$. The marginal
posterior mean (SD) of the shrinkage coefficient $\lambda$ was $5.2$ $(.00)$.
A leave-one-out cross validation analysis indicated that this QIL-based ERG
model fit the Florentine data reasonably well, with 5-number summaries of log
conditional predictive ordinate (log CPO)\ values of
$(-1.6,-1.4,7.7,16.8,16.8)$ over the $n=240=16(15)$ observations
\citep[][p.511]{GelfandDey94}%
. Another analysis of the same data using the QIL-based logit model with LASSO
prior, this time incorporating five covariates, including both original
covariates, their squares, and their two-way interaction, produced marginal
posterior standard deviations with range $(.04,.15)$, and a small CPO improvement.

In contrast, using the Bayesian logit model with the exact likelihood, with
posterior distribution estimated using $10^{5}$ adaptive Metropolis sampling
iterations, the two-covariate model obtained marginal posterior means
$\overline{\boldsymbol{\beta}}=(-1.64,-.01)^{\intercal}$ and marginal
posterior standard deviations $(.29,.15)$, with marginal posterior mean (SD)
for $\lambda$ given by $5.2$ $(.17)$. The five-covariate model resulted in
marginal posterior standard deviations with range $(.09,.62)$ and no CPO
improvement. In summary, the QIL-based approach produced much smaller marginal
posterior uncertainties (standard deviations) for the regression coefficients.

\subsection{Skew-Normal Model for Multivariate iid Data}

Consider data $\mathcal{Y}_{n}=\{\mathbf{y}_{i}\}_{i=1}^{n}$ consisting of
$p$-variate observations with zero means (resp.). For such a data set, the
multivariate skew normal distribution model is defined by the likelihood:%
\begin{equation}
f_{\boldsymbol{\theta}}(\mathcal{Y}_{n})\propto\left.  \exp[-\tfrac{1}{2}%
{\textstyle\sum\limits_{i=1}^{n}}
\mathbf{y}_{i}^{\intercal}\boldsymbol{\Omega}\mathbf{y}_{i}]%
{\textstyle\prod\limits_{i=1}^{n}}
\Phi(\boldsymbol{\alpha}^{\intercal}\mathbf{y}_{i})\right/
|\boldsymbol{\Omega}^{-1}|^{n/2}, \label{skewNormLike}%
\end{equation}
with inverse covariance matrix $\mathbf{\Sigma}^{-1}=\boldsymbol{\Omega
}=(\omega_{jk})_{p\times p}$ and shape parameters $\boldsymbol{\alpha}\in%
\mathbb{R}
^{p}$
\citep{AzzaliniCapitanio99}%
. Recall that the off-diagonal elements of $\boldsymbol{\Omega}$ provide the
quantities $-\omega_{jk}/\sqrt{\omega_{jj}\omega_{kk}}$ that give the partial
correlations between the variable pairs $(Y_{j},Y_{k})$; and that the diagonal
elements of this matrix provide the quantities $1/\omega_{jj}$ that give the
partial variances of the variables $Y_{j}$. Before analyzing any data set
using this model, the observations for each of the $p$ variables can always be
rescaled to have sample mean zero and variance 1, as done throughout this subsection.

For Bayesian inference, the skew normal likelihood (\ref{skewNormLike}) is
difficult to work with\
\citep{LiseoParisi13}%
. In contrast, the QIL\ for this multivariate skew normal model can be
directly specified, to provide posterior inferences of the inverse-covariance
parameters $\boldsymbol{\theta}=\boldsymbol{\Omega}$, while treating
$\boldsymbol{\alpha}$ (and $\boldsymbol{\mu}$) as ignorable nuisance
parameters to simplify such inferences. Specifically, following the
formulation of QIL\ for multivariate iid data ($\S 2.2$), the QIL\ for this
model is specified by the $d$ quantiles $\widehat{\mathbf{q}}_{n,d}$ of the
sample (transformed)\ Mahalanobis depths $D_{R}(\mathbf{y}_{i};\boldsymbol{\mu
}_{\boldsymbol{\theta}},\mathbf{\Sigma}_{\boldsymbol{\theta}})=1-\chi_{p}%
^{2}(M_{\boldsymbol{\theta}}(\mathbf{y}_{i}))$, for $i=1,\ldots,n$, given
$\boldsymbol{\theta}=\boldsymbol{\Omega}$ with $M_{\boldsymbol{\theta}%
}(\mathbf{y}_{i})=\mathbf{y}_{i}^{\intercal}\boldsymbol{\Omega}\mathbf{y}_{i}%
$; and $\mathbf{q}_{\boldsymbol{\theta},d}$ is formed by the corresponding $d$
\textrm{Uniform}$(0,1)$ quantiles; with the quantile vectors
$\widehat{\mathbf{q}}_{n,d}$ and $\mathbf{q}_{\boldsymbol{\theta},d}$ each
defined on $d$ cdf probabilities $\boldsymbol{\lambda}=(\lambda_{d}=\tfrac
{j}{n+1})_{j=1}^{d}$.

Throughout, each data set analyzed using this multivariate skew normal model
was based on QIL using $d(0)=n$, and based on an informative prior
distribution $\boldsymbol{\Omega}\sim\mathrm{Wishart}(\tfrac{1}{p-1}%
\mathbf{I}_{p},p-1)$ for the inverse-covariance matrix parameters
$\boldsymbol{\Omega}=(\omega_{jk})_{p\times p}$. According to $10^{5}$ random
samples from this Wishart prior distribution, the partial correlation
$-\omega_{jk}/\sqrt{\omega_{jj}\omega_{kk}}$ between each variable pair
$(Y_{j},Y_{k})$ has a symmetric prior distribution with median 0, prior
interquartile (50\%)\ range $\pm.24$, and prior 99.9\%\ range $\pm.84$, for
$1\leq j<k\leq p$. Also, the partial variance $1/\omega_{jj}$ of each variable
$Y_{j}$ has a prior distribution with median 1.08, prior interquartile
(50\%)\ range $(.79,1.53)$, and prior 99.9\%\ range $(.32,7.90)$, for $1\leq
j\leq p$.

An extensive simulation study was conducted to evaluate the QIL-based
posterior inference of the multivariate skew normal model, based on the
implementation of the VIS algorithm. Each simulated data set was formed by
generating $n$ iid samples from the zero-mean multivariate normal distribution
with a sparse $10\times10$ correlation matrix, where $35$ of $45$
randomly-selected correlation parameters set to zero, and where the $10$ other
correlation parameters obtained from \textrm{Uniform}$(-1,1)$ draws. Also,
relatively small sample sizes ($n=20,$ $40,$ or $60$) were considered for the
simulation study to ensure large $p/n$. For each of the three sample size
conditions, 100 replications of data sets were made.

From each simulated data set, the posterior distribution of the QIL-based
Bayesian multivariate skew normal model was estimated using $10^{5}$ samples
generated by the VIS algorithm ($\S 2.3.2$). Then, RMSE\ was calculated
separately for entries corresponding to true non-zero correlations, for
entries corresponding to true zero correlations, and for the 10 diagonal
entries. For each of the three groups of entries and each sample size
condition, RMSE\ was taken as the\ root of the QIL importance weighted average
of the squared difference between the VIS samples of the relevant covariance
matrix parameters and the true data-generating covariance parameters,
averaging over all parameters, over all $10^{5}$ VIS samples used for each
simulated data set, and over the 100 simulated data sets.

The results of the simulation study are given in Table 4. The results for
RMSE, for the effective sample size (ESS) and\ computation time of the VIS for
each simulated data set, were found to be acceptable on average for each of
the three sample size conditions. The VIS algorithm provides a fast and useful
algorithm for estimating the posterior distribution of the multivariate skew
normal model, based on the QIL.

\begin{center}%
\footnotesize
\begin{tabular}
[c]{cccccc}%
\multicolumn{6}{l}{\textbf{Table 4.} For the multivariate skew-Normal model
based on the QIL,}\\
\multicolumn{6}{l}{and $p=10$ variables, the mean (SD) for RMSE, ESS, and
computation time}\\
\multicolumn{6}{l}{over $100$ replications.}\\\hline
& RMSE & RMSE & RMSE &  & Computation\\
& for zero & nonzero & diagonal &  & Time\\
$n$ & correlations & correlations & entries & ESS, mean (SD) & Mean
(s.d.)\\\hline
$20$ & $.68$ $(.09)$ & $.36$ $(.03)$ & $.47$ $(.00)$ & $76\,196$ $(1\,395)$ &
$24.6$ secs $(1.1)$\\
$40$ & $.67$ $(.08)$ & $.36$ $(.03)$ & $.47$ $(.00)$ & $71\,049$ $(2\,510)$ &
$29.3$ secs $(1.3)$\\
$60$ & $.68$ $(.07)$ & $.35$ $(.02)$ & $.47$ $(.00)$ & $68\,567$ $(1\,728)$ &
$32.9$ secs $(2.0)$\\\hline
\multicolumn{6}{c}{}%
\end{tabular}%
\normalsize

\end{center}

Finally, we consider the multivariate analysis of the following real data set.
The \texttt{Breast Cancer} data set contains $n=116$ observations of $p=10$
variables of Age (in years), BMI (kg/m2), Glucose (mg/dL), Insulin (%
$\mu$%
U/mL), HOMA, Leptin (ng/mL), Adiponectin (%
$\mu$%
g/mL), Resistin (ng/mL), MCP-1(pg/dL), and breast cancer indicator (1 =
Healthy, 2 = Patients)
\citep{PatricioEtAl18}%
. This data set can be obtained from
\url{https://archive.ics.uci.edu/ml/datasets/Breast+Cancer+Coimbra}%
. We illustrate the QIL-based approach to the multivariate skew normal model,
for the Bayesian analysis of the \texttt{Breast Cancer} data, based on
$10^{5}$ VIS samples used to estimate the posterior distribution. This
analysis took $45$ seconds to complete. Appendix Figure A3 presents the
marginal posterior means and standard deviations of the partial correlations,
and for the partial variances (diagonals elements), of the inverse covariance
matrix $\boldsymbol{\Omega}$.

\subsection{Hierarchical Wallenius Model for Multivariate non-iid Data}

Consider a set of $n$ multivariate data observations $\mathcal{Y}%
_{n}=\{\mathbf{y}_{i}\}_{i=1}^{n}$, where in each observation $\mathbf{y}%
_{i}=(y_{i,1},\ldots,y_{i,c})^{\intercal}$, $y_{i,j}\in\{0,1,\ldots,m_{j}\}$
is the number of objects selected in category $j$ out of $m_{j}$ total
objects, for all $c$ mutually-exclusive categories $k=1,\ldots,c$, with $N=%
{\textstyle\sum\nolimits_{j=1}^{c}}
m_{j}$.

If the data set $\mathcal{Y}_{n}$ consists of $n$ iid multivariate samples,
then these data can be described by a Wallenius distribution model, defined by
the likelihood:%
\begin{equation}
f_{\boldsymbol{\theta}}(\mathcal{Y}_{n})\propto%
{\textstyle\prod\limits_{i=1}^{n}}
{\displaystyle\int\nolimits_{0}^{1}}
{\textstyle\prod\limits_{j=1}^{c}}
\left(  1-t_{i}^{\theta_{j}/\kappa_{i,j}}\right)  ^{y_{i,j}}\mathrm{d}%
t_{i},\ \label{wallLike}%
\end{equation}
with weight parameters $\boldsymbol{\theta}=(\theta_{1},\ldots,\theta
_{c})^{\intercal}$ which are identifiable by the constraints $0\leq\theta
_{j}\leq1$ and $%
{\textstyle\sum\nolimits_{j=1}^{c}}
\theta_{j}=1$
\citep{GrazianEtAl18}%
, and $\kappa_{i,j}=%
{\textstyle\sum\nolimits_{j=1}^{c}}
\theta_{j}(m_{j}-y_{i,j})$
\citep{Wallenius63,Chesson76}%
. The Wallenius likelihood (\ref{wallLike})\ is computationally costly to
evaluate when the sample size $n$ is large, due to its $n$ intractable
integral terms. For posterior distribution inference of the Bayesian Wallenius
iid model, previous research addressed this problem using an ABC method
\citep{GrazianEtAl18}%
.

The QIL can also provide a tractable surrogate for the likelihood
(\ref{wallLike})\ of the Wallenius iid model. Specifically, given that the
Wallenius distribution means $\boldsymbol{\mu}_{\boldsymbol{\theta},i}%
=(\mu_{\boldsymbol{\theta},i,j})_{j=1}^{c}$ and variances $(\sigma
_{\boldsymbol{\theta},i,j}^{2})_{j=1}^{c}$ can be directly calculated given
the model parameters $\boldsymbol{\theta}$
\citep{Fog08calc,Fog15}%
, following the QIL\ formulation for multivariate iid data ($\S 2.2$), it can
be easily shown that the QIL\ $f_{\boldsymbol{\theta}}^{\mathrm{Q}%
}(\mathcal{Y}_{n})$ for the Wallenius model can be obtained from the pivotal
quantities $t_{\boldsymbol{\theta}}(\mathbf{y}_{i})=2(D_{R}(\mathbf{y}%
_{i};\boldsymbol{\mu}_{\boldsymbol{\theta}},\mathbf{\Sigma}%
_{\boldsymbol{\theta}})-1/2)^{2},$ for $i=1,\ldots,n$, using the QIL
formulation for non-iid data, with $D_{R}(\mathbf{y}_{i};\boldsymbol{\mu
}_{\boldsymbol{\theta}},\mathbf{\Sigma}_{\boldsymbol{\theta}})=1-\chi_{p}%
^{2}(M_{\boldsymbol{\theta}}(\mathbf{y}_{i}))$ and $M_{\boldsymbol{\theta}%
}(\mathbf{y}_{i})=%
{\textstyle\sum\nolimits_{j=1}^{c}}
(y_{i,j}-\mu_{\boldsymbol{\theta},i,j})^{2}/\sigma_{\boldsymbol{\theta}%
,i,j}^{2}$. Throughout, we assume $d(0)=n$ for the QIL.

We now illustrate the QIL for the iid Wallenius model through the analysis of
\texttt{Activities} data, obtained from the 2015 Program of International
Student Assessment
\citep[see][]{OECD17}%
, via\newline%
\url{https://nces.ed.gov/pubsearch/getpubcats.asp?sid=098}%
. This data set, shown in Table 5, contains the individual preferences of 56
secondary school students to engage in before- or after-school extracurricular
activities, who were around 15 years of age. Each student was a member of one
of four United States secondary schools that did not offer any extracurricular
activities for students.

In the \texttt{Activities} data set, each student indicated whether they did
any of 11 items in 6 mutually exclusive categories, before and/or after school
during their most recent school day. These categories are: (1) Eat:\ Ate
breakfast or dinner; (2) StudyRead: Study for school or homework, or Read a
book/newspaper/magazine; (3)\ FriendsPlay:\ Internet/Chat/Social networks
(e.g., Facebook, Twitter), Met with friends or talk to friends on the phone,
Watch TV/DVD/Video, or Play video-games; (4) TalkParents:\ Talk to your
parents; (5)\ Work in the household or take care of other family members, or
Work for pay; and (6): Exercise or practice a sport. For the six categories of
11 items, the total number of items are respectively given by $(m_{1}%
,\ldots,m_{6})=(2,4,8,2,4,2)$ with $N=22$. For each person $i$, the
observation vector is given by $\mathbf{y}_{i}=(y_{i,1},\ldots,y_{i,6}%
)^{\intercal}$, where $y_{i,j}\in\{0,1,\ldots,m_{j}\}$ is the number of
activities performed in category $j$, for $j=1,\ldots,c=6$.

\begin{center}%
\footnotesize
\begin{tabular}
[c]{ccc}%
\multicolumn{3}{l}{\textbf{Table 5.} \texttt{Activities} data set.}\\\hline
Category (color) & $m_{j}$ & Observations $\mathbf{y}_{i}$, $i=1,\ldots,n$,
for each of $n=56$ students (columns).\\\hline
Eat & $2$ & 02211221021122111222222121222222122122122222222222222222\\
Study or Read & $4$ &
40124234232244244424204331442434221444220130122434342113\\
Friends or Play & $8$ &
85884446444581287756487565486666366664655674467546546756\\
Talk w/ Parents & $2$ &
20222212222222222222122222222222122122020222122222222222\\
Work & $4$ & 20424122010100041232142240342210041022134424000202100211\\
Exercise & $2$ &
02222112020021121222121121220202020022220222122012111111\\\hline
\end{tabular}%
\normalsize

\end{center}

The \texttt{Activities} data set was analyzed using the Bayesian Wallenius
model, based on QIL\ with $d(0)=n$, and a Dirichlet prior pdf for the choice
weight parameters, given by $\pi(\boldsymbol{\theta})=\mathrm{dirichlet}%
(\boldsymbol{\theta}\mid1,\ldots,1)$. The posterior distribution of this model
was estimated using $10^{5}$ sampling iterations from the adaptive Metropolis
algorithm ($\S 2.3.2$) with $\boldsymbol{\theta}$ starting values given by
$(1/6,...,1/6)$. The analysis completed in $26.3$ minutes. The resulting
marginal posterior distribution estimates of $\boldsymbol{\theta}%
=(\theta,\ldots,\theta_{6})^{\intercal}$ are presented in Appendix Figure A4 (top).

To evaluate the accuracy of the QIL-based inferences of the Bayesian Wallenius
model from the \texttt{Activities} data, a new data set $\{\mathbf{y}%
_{i}\}_{i=1}^{n}$ was simulated by taking $n=56$ iid samples from the
Wallenius distribution model, with true choice weight parameters
$\boldsymbol{\theta}$ given by the posterior mean estimates $\overline{\theta
}=(.10,.17,.12,.29,.14,.18)$ obtained from the previous \texttt{Activities}
data analysis, along with $(m_{1},\ldots,m_{6})=(2,4,8,2,4,2)$ as before.
After this data set was simulated, the posterior distribution
$\boldsymbol{\theta}$ for this model was estimated using $10^{5}$ sampling
iterations from the adaptive Metropolis algorithm, using $(1/6,...,1/6)$ as
the starting values for $\boldsymbol{\theta}$. This analysis completed in
$29.3$ minutes. We found that RMSE =$.03$, indicating high accuracy in the
estimation of $\boldsymbol{\theta}$. This RMSE was based on averaging squared
error estimation loss over the $c$ weight parameters and over the $10^{5}$
Metropolis posterior samples.

It can be argued that the Bayesian Wallenius model considered thus far makes
the overly-restrictive assumption that weight vector parameters
$\boldsymbol{\theta}=(\theta_{1},\ldots,\theta_{c})^{\intercal}$ are shared by
all $n$ persons in the data, which implies that the $n$ choice data
observations $\mathcal{Y}_{n}=\{\mathbf{y}_{i}\}_{i=1}^{n}$ are iid. We can
relax this iid assumption by the considering of the new, Bayesian hierarchical
Wallenius model, suitable for the analysis of non-iid choice data
$\mathcal{Y}_{n}=\{\mathbf{y}_{i}\}_{i=1}^{n}$. This new model is based on the
original Wallenius likelihood (\ref{wallLike}), after replacing
$\boldsymbol{\theta}=(\theta_{1},\ldots,\theta_{c})^{\intercal}$ with the
choice weight parameter $\boldsymbol{\theta}_{i}=(\theta_{i,1},\ldots
,\theta_{i,c})^{\intercal}$, defined for each individual $i=1,\ldots,n$. Now,
each individual $i$ has her own personal choice weight parameters
$\boldsymbol{\theta}_{i}$. The choice weight parameters $\{\boldsymbol{\theta
}_{i}\}_{i=1}^{n}$ parameters over all $n$ persons are then assigned a prior distribution.

To illustrate, we analyze the \texttt{Activities} data using the Bayesian
hierarchical Wallenius model, with the logit of the choice weight parameters
assigned the 5-variate normal prior distribution, with $\{(\mathrm{log}%
(\theta_{i,1}/\theta_{i,6}),\ldots,\mathrm{log}(\theta_{i,5}/\theta
_{i,6}))\}_{i=1}^{n=56}\overset{\text{iid}}{\sim}\mathrm{N}_{c-1}%
(\mathbf{0},\mathbf{I}_{5})$. The QIL\ for the hierarchical Wallenius model is
specified in the same manner as done before for the Bayesian Wallenius model
for iid choice data, after replacing $\boldsymbol{\theta}$ with the personal
choice weight parameters $\{\boldsymbol{\theta}_{i}\}_{i=1}^{n}$. It turns out
that performing a QIL-based analysis\ using the Bayesian hierarchical
Wallenius model for non-iid choice data, is as easy as performing QIL-based
analysis of the Bayesian Wallenius model for iid data.

For the hierarchical Wallenius model under QIL, the posterior distribution of
the choice weight parameters for each person was estimated by $10^{5}$
sampling iteration run of a random-walk Metropolis algorithm for the logit of
the choice weight parameters, which completed in $32$ minutes. Here, the
proposal variance .176 was selected to approximate a .234 acceptance rate over
iterations, on average over the 56 persons.

Appendix Figure A4 (middle and bottom panels)\ presents the estimates of the
marginal posterior means and standard deviations of the 6 choice weight
parameters, for the 56 persons, respectively. These results show
between-person differences in the marginal posterior distribution of the
weight parameters, thus providing evidence of between-person differences in
choice behavior, which is not surprising.

\section{Conclusions}

We have introduced a general framework for likelihood-free Bayesian inference,
which employs a tractable QIL as a surrogate to the possibly-intractable exact
likelihood of the given Bayesian model. The QIL\ is defined by an asymptotic
multivariate normal pdf of the quantiles implied by the likelihood given
parameters, under reasonable assumptions for large data sets. An appealing
feature of the QIL\ approach is its generality and applicability. We have
shown that the QIL\ can be constructed in an automatic manner for a wide range
of Bayesian intractable-likelihood models for univariate or multivariate iid
or non-iid data. QIL\ can also lead to accurate posterior inferences with
improved or competitive computational speed in posterior distribution
estimation, compared to previous likelihood-free methods. Unlike the previous
methods, QIL can be easier to implement because it is automatically specified
by the Mahalanobis distance measure and multivariate normal kernel density for
quantiles, according to asymptotic theory; while quantiles provide fundamental
summary statistics of distributions. Also, QIL\ avoids the computationally
costly tasks of synthetic data sampling, point estimation, and the selection
of many tuning parameters.

We believe that the QIL\ offers many exciting possibilities for statistical
analysis involving intractable likelihoods. In a similar spirit to
distribution testing, each model parameter $\boldsymbol{\theta}$ that is
hypothesized (proposed)\ in a posterior distribution estimation algorithm (via
sampling or optimization)\ provides a test of the fit of that parameter to the
data, with the plausibility of $\boldsymbol{\theta}$ measured by the pdf of
the distribution of the quantile-based statistic under the null hypothesis.
This simple but general statistical idea is what makes the QIL\ applicable to
a wide range of Bayesian models. We expect that this property is important in
identifying new future application areas and for the development of new
related methodologies, which may employ new quantile-based distribution test
statistics that can be computed from the given intractable likelihood.

\bigskip

\noindent\textbf{Acknowledgements}

George Karabatsos (gkarabatsos1@gmail.com) is the lead corresponding author,
University of Illinois-Chicago, 60607. Karabatsos' and Leisen's research was
respectively supported in part by NSF grant SES-1156372, and by the European
Community's Seventh Framework Programme [FP7/2007-2013] under grant agreement
no. 630677.\newline

\noindent\noindent
\bibliographystyle{ba}
\bibliography{Karabatsos}

\end{document}